\documentclass[preprint2,twocolumn,times,tighten,twocolappendix]{aastex63}

\usepackage{color}
\usepackage{enumitem}
\usepackage{hyperref}
\usepackage{verbatim}
\usepackage{amsmath,amstext,mathrsfs}
\usepackage[all]{hypcap} 
\usepackage{afterpage}    
\usepackage{marginnote}
\usepackage{makecell}
\usepackage{subfigure}
\usepackage{multirow}

\shorttitle{Synergy of PDFs and the VGT}
\shortauthors{Hu \& Lazarian}
\graphicspath{{./}{figures/}}

\begin{document}
	
\title{A Synergy of the Velocity Gradients Technique and the Probability Density Functions for Identifying Gravitational Collapse in Self-Absorbing Media }

\email{yue.hu@wisc.edu; alazarian@facstaff.wisc.edu}
	
\author[0000-0002-8455-0805]{Yue Hu}
\affiliation{Department of Physics, University of Wisconsin-Madison, Madison, WI 53706, USA}
\affiliation{Department of Astronomy, University of Wisconsin-Madison, Madison, WI 53706, USA}
\author{A. Lazarian}
\affiliation{Department of Astronomy, University of Wisconsin-Madison, Madison, WI 53706, USA}
\affil{Center for Computation Astrophysics, Flatiron Institute, 162 5th Ave, New York, NY 10010}

\begin{abstract}
The Velocity Gradients Technique (VGT) and the Probability Density Functions (PDFs) of mass density are tools to study turbulence, magnetic fields, and self-gravity in molecular clouds. However, self-absorption can significantly make the observed intensity different from the column density structures. In this work, we study the effects of self-absorption on the VGT and the intensity PDFs utilizing three synthetic emission lines of CO isotopologs $^{12}$CO (1-0), $^{13}$CO (1-0), and C$^{18}$O (1-0). We confirm that the performance of VGT is insensitive to the radiative transfer effect. We numerically show the possibility of constructing 3D magnetic fields tomography through VGT. We find that the intensity PDFs change their shape from the pure log-normal to a distribution that exhibits a power-law tail depending on the optical depth for supersonic turbulence. We conclude the change of CO isotopologs' intensity PDFs can be independent of self-gravity, which makes the intensity PDFs less reliable in identifying gravitational collapsing regions. We compute the intensity PDFs for a star-forming region NGC 1333 and find the change of intensity PDFs in observation agrees with our numerical results. The synergy of VGT and the column density PDFs confirms that the self-gravitating gas occupies a large volume in NGC 1333. 
\end{abstract}
	

	\section{Introduction}
	\label{sec:intro}
	Magnetic and turbulent effects are considered to be the crucial agents affecting the dynamics of the star-forming process in molecular clouds, in combination with gas self-gravity, at all physical scales and throughout different evolutionary stages \citep{1965ApJ...142..584P,1979cmft.book.....P,1966ApJ...146..480J,MO07,2011Natur.479..499L,2013ApJ...768..159H,2014ApJ...783...91C,Aetal15}. To understand this complex interplay, it is essential to explore the properties of turbulence, trace the magnetic fields, and identify the transition regions where the gravity takes over, and the collapse occurs \citep{1977ApJ...214..488S,1992pavi.book.....S,1994ApJ...429..781S,1998ApJ...498..541K,1998ARA&A..36..189K,2011ApJ...730...40P,2012ApJ...761..156F,2020MNRAS.491.4310T}.
	
	The probability density functions (PDFs) of the column mass density is a statistical tool to get an insight into turbulence and self-gravity in molecular clouds. The column density PDFs are believed to be log-normal in non-self-gravitating isothermal supersonic turbulence \citep{1995ApJ...441..702V,2000ApJ...535..869K,2008ApJ...680.1083R,2011ApJ...727L..20K,2012ApJ...750...13C,2017ApJ...840...48P,2018ApJ...863..118B}:
	\begin{equation}
		P_{N}(s)=\frac{1}{\sqrt{2\pi\sigma_s^2}}e^{-\frac{(s-s_0)^2}{2\sigma_s^2}}
	\end{equation}
	where s$=ln(\rho/\rho_0)$ is the logarithmic density and $\sigma_s$ is the standard deviation of the log-normal, while $\rho_0$ and $s_0$ denote the mean density and mean logarithmic density. The log-normal PDFs of column density data are used to predict the sonic Mach number \citep{2010ApJ...708.1204B,2011ApJ...727L..21P}, the star formation rate \citep{2005ApJ...630..250K,2012ApJ...761..156F}, and the initial mass function \citep{2008ApJ...684..395H,2011ApJ...743L..29H} in both diffuse and dense ISM medium. In addition, in the presence of gravitational collapsing, the column density PDFs are believed to be shaped into a log-normal ($P_{N}$) format for low-density gas and a power-law ($P_{L}$) format for self-gravitating gas \citep{1995ApJ...441..702V,2008ApJ...680.1083R,2012ApJ...750...13C,2018ApJ...863..118B,2019MNRAS.482.5233K}:
	\begin{equation}
		P_{L}(s)\propto e^{k s},s>s_t
	\end{equation}
	where s$_t = ln(\rho_t/\rho_0 )=(-k-\frac{1}{2})\sigma_s^2$ is the logarithm of the normalized transitional density between the $P_{N}$ and $P_{L}$. The transition from log-normal to power-law column density PDFs reveals the density threshold, above which the gas becomes self-gravitating. However, molecular clouds are not optically thin objects. The effects of self-absorption that vary with both gas density and abundance of molecular species can significantly changed the observed intensity structures \citep{2001ApJ...546..980O,2004A&A...416..191T}. The intensity distribution of molecular species does not represent the true density distribution in molecular clouds \citep{2013ApJ...771..122B,2016MNRAS.460...82S}. It therefore naturally raises a question: how does the effect of radiative transfer affect the corresponding intensity PDFs?
	
	Nevertheless, the Velocity Gradients Technique (VGT) provides an alternative way to solve this problem. VGT is developed as a new technique to study the magnetic fields in ISM \citep{2017ApJ...835...41G, YL17a, LY18a,PCA}, based on the advanced understanding of MHD turbulence \citep{GS95,LV99}. Due to anisotropic properties of turbulent eddies, i.e., the eddies are elongating along their local magnetic fields, their velocity gradients are perpendicular to the magnetic fields \citep{2000ApJ...539..273C,2001ApJ...554.1175M,2002ApJ...564..291C}. The magnetic field direction can, therefore, be inferred from the velocity gradients. One of the most critical properties of velocity gradients is that the gradients flip their orientation by 90$^\circ$, i.e., get changed from perpendicular to magnetic fields to parallel with magnetic fields, in the cases of gravitational collapsing \citep{YL17b,survey,Hu20}. Therefore, the reaction of gradients with respect to self-gravity provides an alternative way to identify gravitational collapsing regions. 
	
	The first attempt to study mangetic field direction via velocity gradients was made by \cite{2017ApJ...835...41G}. The sub-block averaging method \citep{YL17a} and the principal component analysis \citep{PCA} futher boost the VGT to higher accuracy.
	Its robustness of tracing the magnetic fields morphology has been successful tested in transparent diffuse gas \citep{YL17a,2019ApJ...874...25G,2020RNAAS...4..105H}.
	The effects of radiative transfer, however, can make a big difference for observations of the measured intensities. This difference benefits the VGT in tracing the magnetic fields and identifying gravitational collapsing regions. For instance, \citet{velac} firstly showed the possibility of constructing the 3D magnetic fields model in molecular clouds utilizing different emission lines. As an example, the velocity gradient can resolve the plane-of-the-sky (POS) magnetic field over three different density ranges using $^{12}$CO, $^{13}$CO, and C$^{18}$O. We can, therefore, expect that VGT can reveal the density range in which the collapsing occurs. In this work, utilizing the SPARX radiative transfer code \citep{2019ApJ...873...16H}, we generate three synthetic emission lines of CO isotopologs, i.e., $\rm ^{12}CO$(1-0), $\rm ^{13}CO$(1-0), $\rm C^{18}O$(1-0). We then study how the PDFs of measured intensitiesa, as well as the VGT, are affected by the radiative transfer. We numerically investigate the effects of molecular abundances, optical depths, and molecular emission lines. For observations, we test the results in the low-mass star-forming region NGC 1333.
	
	The paper is organized as follows. In \S~\ref{Sec.data}, we give the details of the numerical simulation used in this work. In \S~\ref{sec:results} and \S~\ref{sec:obs}, we numerically and observationally compare the VGT and the PDFs. We discuss the physical implication of the PDFs and VGT in self-absorbing and self-gravitating media in \S~\ref{sec.dis} and give our conclusion in \S~\ref{sec:con}.
	
	\begin{table}
		\centering
		\label{tab:sim}
		\begin{tabular}{ | c | c | c | c | c | c | c| }
			\hline
			Model & $\rm M_S$ & $\rm M_A$ & Snapshots & $\tau_{12}$ & $\tau_{13}$ & $\tau_{18}$\\ \hline \hline
			$\rm M_A$0.2 & 7.31 & 0.22 & 0 Myr & 181.594 & 3.467 & 0.295\\ 
			$\rm M_A$0.4 & 6.10 & 0.42 & 0 - 0.8 Myr  & 49.515 & 0.945 & 0.080\\
			$\rm M_A$0.6 & 6.47 & 0.61 & 0 Myr & 21.650 & 0.413 & 0.035\\  
			$\rm M_A$0.8 & 6.14 & 0.82 & 0 Myr & 12.597 & 0.240 & 0.020\\  
			$\rm M_A$1.0 & 6.03 & 1.01 & 0 Myr & 8.011 & 0.152 & 0.013\\
			$\rm M_A$1.2 & 6.08 & 1.19 & 0 Myr & 5.655 & 0.107 & 0.009\\
			$\rm M_A$1.4 & 6.24 & 1.38 & 0 Myr & 4.000 & 0.076 & 0.007\\
			$\rm M_A$1.6 & 5.94 & 1.55 & 0 Myr & 3.242 & 0.061 & 0.005\\
			$\rm M_A$1.8 & 5.80 & 1.67 & 0 Myr & 2.630 & 0.050 & 0.004\\
			$\rm M_A$2.0 & 5.55 & 1.71 & 0 Myr & 2.180 & 0.041 & 0.004\\
			\hline
		\end{tabular}
		\caption{Description of our MHD simulations. $\rm M_S$ and $\rm M_A$ are the instantaneous values at the snapshots 0 Myr, after which the self-gravity is introduced. The lowest-transition J = 1-0 is considered for all CO isotopologs. The fractional abundances of $^{12}$CO, $^{13}$CO, and C$^{18}$O are set as $1\times10^{-4}$, $2\times10^{-6}$, and $1.7\times10^{-7}$, respectively. $\tau_{12}$, $\tau_{13}$, and $\tau_{18}$ represent the mean optical depth of $^{12}$CO, $^{13}$CO, and C$^{18}$O, respectively.}
	\end{table}
	

	\section{Numerical simulation}
	\label{Sec.data}
	We simulate numerical 3D MHD simulations through the ZEUS-MP/3D code \citep{2006ApJS..165..188H}, which solves the ideal MHD equations in a periodic box:
	\begin{equation}
		\begin{aligned}
			&\partial\rho/\partial t +\nabla\cdot(\rho\Vec{v})=0\\
			&\partial(\rho\Vec{v})/\partial t+\nabla\cdot[\rho\Vec{v}\Vec{v}+(p+\frac{B^2}{8\pi})\Vec{I}-\frac{\Vec{B}\Vec{B}}{4\pi}]=f\\
			&\partial\Vec{B}/\partial t-\nabla\times(\Vec{v}\times\Vec{B})=0
		\end{aligned}
	\end{equation}
	where $f$ is a random large-scale driving force, $\rho$ is the density, $\Vec{v}$ is the velocity, and $\Vec{B}$ is the magnetic field. We also consider a zero-divergence condition $\nabla \cdot\Vec{B}=0$, and an isothermal equation of state $p = c_s^2\rho$, where $p$ is the gas pressure. We assume a single fluid, operator-split, staggered grid MHD in Eulerian frame. Periodic boundary conditions and solenoidal turbulence injections at wave scale $k$ equal to 2 are applied in our simulations. The simulated interstellar clouds are isothermal with temperature T = 10.0 K and sound speed $c_s$ = 187 m/s. To probe the relative importance of gravity and thermal pressure forces, we consider cloud with the size L = 10 pc and vary the initial density $\rho_0$. We vary the value of $\rho_0$, magnetic field strength $B$ to stipulate different physical environments. The sound crossing time $t_v = L/c_s$ is $\sim 52.0$ Myr, which is fixed owing to the isothermal equation of state.
	
	The physical conditions are characterized by the Alfv\'{e}nic Mach numbers M$_{A}=\sqrt{4\pi\rho_0}v_{L}/B$, and the Sonic Mach number M$_{S}=v_{L}/c_{s}$, where $v_{L}$ is the injection velocity and $v_{A}$ is the Alfv\'{e}nic velocity. The supersonic simulations can be divided into two groups corresponding to sub-Alfv\'{e}nic ($\rm M_A<1$) and super-Alfv\'{e}nic ($\rm M_A>1$) turbulence. In the case of $\rm M_{A}<M_{S}$, the cloud is highly magnetized while $\rm M_{A}>M_{S}$ corresponds to the thermal pressure and turbulence dominate the cloud. We list the simulations in Tab.~\ref{tab:sim}, which have been utilized in \citet{YL17b} . In the text and figures, we refer to the corresponding simulation by their model name. 
	
	For simulation $\rm M_A0.4$, we additionally consider the effect of self-gravity. We employ a periodic Fast Fourier Transform Poisson solver for the self-gravitating module and keep driving both turbulence and self-gravity until the evolution time $t\simeq 0.43$ $t_{ff}$, where $t_{ff} \simeq 1.88$ $\rm Myr$ is the free-fall time. The total mass $\rm M_{tot}$ in the simulated cubes $\rm M_A0.4$ is $\rm M_{tot}\simeq8430.785 M_\odot,$ the magnetic Jean mass is $\rm M_{JB}\simeq86.95 M_\odot$, average magnetic field strength is $\rm B\simeq30.68 \mu G$, mass-to-flux ratio is $\Phi\simeq1.11$, and mass density is $\rm \rho\simeq315$ g $\rm cm^{-3}$.
	
	\subsection{Radiative transfer}
	\label{subsec.MHD data}
	In this work, we generate three synthetic emission lines of CO isotopologs, i.e., $^{12}$CO(1-0), $^{13}$CO(1-0), C$^{18}$O(1-0) utilizing the SPARX radiative transfer code \citep{2019ApJ...873...16H}. The SPARX solves radiative transfer equation (RTE) for the finite cells, i.e., the emission from a homogeneous finite element:
	\begin{equation}
		\label{eq.4}
		\frac{dI_\nu}{ds}=-k_\nu I_\nu+\epsilon_\nu
	\end{equation}
	in which $I_\nu$, $k_\nu$, $\epsilon_nu$ is the specific intensity, the absorption coefficient, and the emission coefficient at a given frequency $\nu$, respectively. The SPARX utilizes the Accelerated Lambda Iteration (ALI) to describe the radiative interaction, which reaches spatial preliminary consistency at the first stage then improve the random sampling resolution to the demanding accuracy at the next stage. In general, ALI differentiates the intensity inside a cell into the contribution from the internal intensity of the cell $ J_{int}$ and the contribution from the external cell $J_{ext}$ to reduce the computational consumption of ray-tracing. Between the iteration, ALI samples $J_{ext}$ once then performs a detailed balance calculation to make $J_{int}$ and population self-consistent. The equation of statistical equilibrium about the molecular levels considers molecular self-emission, stimulated emission, and the collision with the gas particles. The information about molecular gas density and velocity is extracted from the MHD data mentioned above. The cube is observed at a distance of 10 kpc with a velocity resolution of 0.02 km $\rm s^{-1}$ and beam width 0.26$''$ .
	
	The fractional abundances of the CO isotopologs $^{12}$CO(1-0), $^{13}$CO(1-0), and C$^{18}$O(1-0) are set as $1\times10^{-4}$, $2\times10^{-6}$, and $1.7\times10^{-7}$, respectively, following \citet{2019ApJ...873...16H}. The $^{12}$CO-to-H$_2$ ratio of $1\times10^{-4}$ comes from the cosmic value of C/H = $3\times10^{-4}$ and the assumption that 15\% of C is in the molecular form. For the abundance of $^{13}$CO, we adopted a $^{13}$CO/$^{12}$CO ratio of 1/69 \citep{1999RPPh...62..143W}. Hence, the $^{13}$CO-to-H$_2$ ratio is approximated to $2\times10^{-6}$. With $^{12}$CO/C$^{18}$O = 500 \citep{Wilson16}, we obtain a C$^{18}$O-to-H$_2$ ratio of $1.7\times10^{-7}$.  When producing the synthetic molecular channel maps, we focus on the lowest-transition J = 1-0 of the CO isotopologs, in which the LTE condition is satisfied. 

	\section{The Velocity Gradients Technique}
	\label{sec.theory}
	\subsection{MHD turbulence theory}
	The Velocity Gradient Technique (VGT) is initially developed as an advanced tool for magnetic field studies \citep{2017ApJ...835...41G,YL17a,LY18a,PCA}. It is rooted in the advanced magnetohydrodynamic (MHD) turbulence theory \citep{GS95} and the turbulent reconnection theory \citep{LV99}. \citet{GS95} firstly proposed that within MHD turbulence, the turbulent eddies are anisotropic. This anisotropy is increasing with the decrease of eddies' scale, exhibiting the scaling relation:
	\begin{equation}
		\label{eq:gs} 
		k_\parallel\propto(k_\bot)^{2/3}
	\end{equation}
	which is known as GS95 anisotropy scaling. Here, $k_\parallel$ and $k_\bot$ wavenumbers perpendicular and parallel to the magnetic field, respectively. However, the GS95 scaling is derived within the mean-field reference frame, in which the anisotropy cannot be observed. This system of reference is also called a global system of reference.
	
	It is crucial that in order to get the correct anisotropy scaling, one should use a different system of reference. This system of reference can be understood on the basis of the theory of turbulent reconnection in \citet{LV99}. There it was shown that Alfv\'{e}nic turbulence could be presented as a collection of eddies mixing up plasmas perpendicular to the direction of the magnetic field percolating the eddies. This magnetic field at the eddy's position is termed the {\it local} magnetic field direction. Naturally, the local direction changes in space following the wandering of the magnetic field in the volume. The local magnetic field's notion is crucial for many physical processes, e.g., cosmic ray propagation (see \citealt{2002PhRvL..89B1102Y}). It is also essential, as we discuss below, for the VGT. 
	
	The anisotropy scaling relation in the {\it local} reference frame are given \citep{LV99}:
	\begin{equation}
		\label{eq.lv99}
		l_\parallel\simeq L_{inj}(\frac{l_\bot}{L_{inj}})^{\frac{2}{3}}M_A^{-4/3}
	\end{equation}
	where $l_\bot$ and $l_\|$ are the perpendicular and parallel size of eddies in respect to the {\it local} magnetic field. In Eq. (\ref{eq.lv99}) the turbulence is assumed subAlfv\'{e}nic with $\rm M_A$ being the Alfv\'{e}n Mach number, i.e., the ratio of the injection velocity $v_{inj}$ to the Alfv\'{e}n speed $v_A$. $L_{inj}$ is the injection scale of turbulence. 
	
	Eq.~\ref{eq.lv99} is derived in the framework of the turbulent reconnection theory. As we mentioned earlier, due to fast reconnection \citet{LV99} demonstrated that the turbulent motions of strongly magnetized fluid present eddies aligned with the {\it local} magnetic field direction. The eddy motions are not constrained as the time scale for the turbulent reconnection coincides with the eddy turnover time.  As a consequence, eddies that are freely mixing magnetic field lines perpendicular to their direction. These hydrodynamic-type motions are possible due to the absence of magnetic back-reaction. In fact, this motion of eddies presents the path of minimal resistance. This channel the energy cascade in an anisotropic way. 
	
	The universal scale-dependent anisotropy of Alfv\'{e}nic turbulence in the local magnetic field reference frame has been demonstrated in numerical simulations \citep{2000ApJ...539..273C,2001ApJ...554.1175M,2002ApJ...564..291C}. This anisotropy indicates that the eddies are aligned with the {\it local} magnetic fields. By detecting this anisotropic direction, we can determine the magnetic field direction at the eddy location. The velocity gradients, which are perpendicular to the eddies, play the role of a detector. Once we obtain the velocity gradients' direction and rotate them by 90$^\circ$, we can map the magnetic field direction in turbulent media. This is the theoretical cornerstone of the VGT.
	
	In particular, the velocity gradient and density gradient scale as \citep{2018arXiv180200024Y}: 
	\begin{equation}
		\label{eq.grad}
		\begin{aligned}       
			\nabla \rho_l&\propto\frac{\rho_{l}}{l_\bot}\simeq\frac{\rho_0}{c}\mathscr{F}^{-1}(|\hat{k}\cdot\hat{\zeta}|)\nabla v_l\\
			\nabla v_l&\propto\frac{v_{l}}{l_\bot}\simeq \frac{v_{inj}}{L_{inj}}(\frac{l_{\perp}}{L_{inj}})^{-\frac{2}{3}}M_A^{\frac{1}{3}}
		\end{aligned}
	\end{equation}
	because the anisotropic relation indicates $l_\bot \ll l_\parallel$. Here $\rho_0$ is the mean density, $\hat{\zeta}$ is the unit vector for the Alfv\'{e}nic mode, fast mode, or slow mode, c is the  propagation speed of corresponding mode. $\mathscr{F}^{-1}$ denotes the inverse Fourier transformation.
	
	\subsection{The VGT in self-absorbing and self-gravitating media}
	The effect of radiative transfer imposes a significant modification on the observed density structures. The specific intensity $I_\nu$ at a given frequency $\nu$ (see Eq.~\ref{eq.4}) can be re-arranged into form: 
	\begin{equation}
		\frac{dI_\nu}{d\tau_\nu}=-I_\nu+S_\nu
	\end{equation}
	where $d\tau_\nu = k_\nu ds$ is the optical depth, and $S_\nu=\epsilon_\nu/k_\nu$ is the source function. $k_\nu$ and $\epsilon_\nu$ are related to the Einstein coefficients $A$, $B$ and the molecular gas density $n(\Vec{r})=X\rho(\Vec{r})$, where $X$ is the abundance and $\rho(\Vec{r})$ is the H$_2$ volume density distributed in real space:
	\begin{equation}
		\begin{aligned}
			k_\nu^{ij}(\Vec{r})&=n_{ij} A_{ij}\phi_\nu(\Vec{r})\\
			\epsilon_\nu^{ij}(\Vec{r})&=(n_jB_{ji}-n_iB_{ij})\phi_\nu(\Vec{r})
		\end{aligned}
	\end{equation}
	where $i$ and $j$ denote the starting and ending energy states of the molecular transition under consideration. The LOS component of velocity $v$ at the position $\Vec{r}=(x,y,z)$ is a sum of the regular gas flow due to Galactic rotation $v_{gal}(\Vec{r})$, the turbulent velocity $\mu(\Vec{r})$ and the residual component due to thermal motions. The Doppler broadening function $\phi_\nu(\Vec{r})$ expressed in terms of velocity can be formed into a Maxwellian distribution of this residual thermal velocity $v-v_{gal}(\Vec{r})-\mu(\Vec{r})$\citep{2004ApJ...616..943L}:
	\begin{equation}
		\label{eq.max}
		\phi_v(\Vec{r})=\frac{1}{\sqrt{2\pi\beta(\Vec{r})}}\exp[-\frac{(v-v_{gal}(\Vec{r})-\mu(\Vec{r}))^2}{2\beta(\Vec{r})^2}]
	\end{equation}
	where $\beta(\Vec{r})=k_BT(\Vec{r})/m$, $m$ being the mass of atoms or molecules. The temperature $T(\Vec{r})$ can vary from point to point if the emitter is not isothermal. By assuming the abundance $X$ and Einstein coefficients, the solution of observed intensity $I_\nu(x,y)$ is:
	\begin{equation}
		\begin{aligned}
			\label{eq.11}
			I_\nu(x,y)&=I_\nu^{e}e^{-\tau_\nu}+S_\nu(1-e^{-\tau_\nu})\\
			&=(I_\nu^{e}-S_\nu)e^{-AX\int_0^s\rho(\Vec{r})\phi_v(\Vec{r})dz}+S_\nu
		\end{aligned}
	\end{equation}
	here $I_\nu^{e}$ is the intensity of external illumination. The integration variable:
	\begin{equation}
		Y_v(x,y,s)=\int_0^s \rho(\Vec{r})\phi_v(\Vec{r})dz
	\end{equation}
	coincidentally has it value equivalent to the density distribution $\rho(x,y,v)$ in PPV coordinates \citep{2004ApJ...616..943L}:
	\begin{equation}
		\rho_s(x,y,v)dv=[\int_0^s\rho(\Vec{r})\phi_v(\Vec{r})dz] dv
	\end{equation}
	which counts the number of molecules along the LOS that have a z-component of velocity in the interval [$v$,$v+dv$]. Eq.~\ref{eq.11} can therefore be expressed as:
	\begin{equation}
		I_\nu(x,y,v)=(I_\nu^{e}-S_\nu)e^{-\alpha\rho_s(x,y,v)}+S_\nu
	\end{equation}
	where $\alpha=AX$. Here we consider an isothermal model for molecular clouds so that the source function $S_v$ and $I_\nu^e$, which here is the radiation from CMB background, are constant for a given molecular specie. The total intensity map $I(x,y)$ integrated along the LOS in PPV space is then:
	\begin{equation}
		\label{eq.16}
		\begin{aligned}
			I(x,y)&=\int\{(I_\nu^e-S_\nu)e^{-\alpha\rho_s(x,y,v)}+S_\nu\} dv\\
			&=\int I_\nu^ee^{-\alpha\rho_s(x,y,v)}dv+\int S_\nu(1-e^{-\alpha\rho_s(x,y,v)})dv
		\end{aligned}
	\end{equation}
	In the case of vanishing absorption and external illumination, the intensity is given by the linear term in the expansion of the exponent in Eq.~\ref{eq.16}:
	\begin{equation}
		I(x,y)=S_\nu\alpha\int \rho_s(x,y,v)dv
	\end{equation}
	and reflects the PPV density of the emitters.
	The gradient of $I(x,y)$ in the POS is then:
	\begin{equation}
		\begin{aligned}
			&\nabla_{pos} I(x,y)=-\alpha(I_\nu^{e}-S_\nu)\int e^{-\alpha\rho_s(x,y,v)}\nabla_{pos} \rho_s(x,y,v)dv\\
			&=-\alpha(I_\nu^{e}-S_\nu)\int e^{-\alpha\rho_s(x,y,v)}dv\int\nabla_{pos}[\rho(\Vec{r})\phi(\Vec{r})]dz
		\end{aligned}
	\end{equation}
	in which the gradient term can be written as:
	\begin{equation}
		\label{eq.18}
		\begin{aligned}
			&\int\nabla_{pos}[\rho(\Vec{r})\phi(\Vec{r})]dz=\\
			&\int\phi[\nabla_{pos}\rho-\frac{\rho}{\beta}(v-v_{gal}-\mu)(\nabla_{pos} v_{gal}+\nabla_{pos}\mu)]dz
		\end{aligned}
	\end{equation}
	$\nabla_{pos} I(x,y)$ therefore includes the contribution from density $\rho$, the regular gas flow due to Galactic rotation $v_{gal}$, and the turbulent velocity $\mu$. Considering the flux freezing condition\footnote{The flux freezing in its classical sense is violated in the presence of turbulent reconnection \citep{LV99} and is substituted by the concept of the stochastic flux freezing \citep{Eyink2012}, which describes the approach of describing the magnetic field structure in the presence of the reconnection diffusion. For small-amplitude sub-Alfvenic turbulence, the diffusion of the magnetic field induced by turbulence is small.}, the magnetic field is expected to follow the radial direction of regular gas flow caused by Galactic rotation. Coincidentally, the gradients term $\nabla_{3D} v_{gal}$, in this case, are still perpendicular to the magnetic field, showing a different scaling relation from Eq.~\ref{eq.grad}. Therefore, the gradients' direction is not changed by the radiative transfer, but only the gradients' amplitude. When the thermal velocity is significant, i.e., $\beta$ is large, the resulting gradient  $\nabla_{pos} I(x,y)$ is dominating by the density's contribution.
	
	The study in \citet{2019MNRAS.483.1287G} numerically confirms the validity of the VGT in $\rm ^{13}CO$ (2-1) medium with different $\rm ^{13}CO$ abundances, densities, and optical depths. \citet{2019ApJ...873...16H} extend the study to both intensity gradient and velocity gradient using three different CO isotopologs, i.e., $\rm ^{12}CO$ (1-0), $\rm ^{13}CO$ (1-0), and $\rm C^{18}O$ (1-0). Their studies showed that VGT is applicable to various molecular lines and different physical conditions, in terms of tracing the magnetic fields' orientation. The observation demonstrations are provided by \citet{survey, velac,2020arXiv200715344A}.

	However, in the presence of gravitational collapse, the self-gravity radically modifies the nature of the turbulent flow.
	When gravity is sub-dominant to magnetic and turbulent energy, the magnetized turbulent eddies are elongated in the direction parallel to the magnetic field surrounding the eddies. As discussed above, the maximum change of the velocity amplitudes (i.e., velocity gradient), is in the direction perpendicular to the local magnetic field, and by rotating the velocity gradient by 90$^\circ$ we can trace the magnetic field. However, in the case of gravitational collapse, the gravitational force produces the most significant acceleration of the plasma in the direction parallel to the magnetic field and the velocity gradients are parallel to the magnetic field. This happens in both strong and weak magnetic field environments as the plasma's infall motions will alter the magnetic field geometry so that it tends to align parallel to the direction of gravitational collapse. Therefore, in strong self-gravitating media, the velocity gradients are expected to change their orientation from orthogonal to magnetic fields to align with magnetic fields \citep{YL17b, survey,Hu20}. This particular reaction of gradients to the self-gravity is developed as a new tool in identifying the gravitational collapsing region through the histogram of gradients' orientation \citep{Hu20}. 
	
	\subsection{The VGT algorithm}
	In observations, velocity information is usually obtained from spectroscopic Position-Position-Velocity (PPV) cubes. \citet{LP00} firstly presents the statistics of the intensity fluctuations in PPV and their relations to the underlying statistics of turbulent velocity and
	density. There, it is shown that the velocity fluctuation can be most prominent in thin velocity slices of PPV cubes. This happens when the velocity channel is thin enough, i.e., the channel width $\Delta v$ satisfies:
	\begin{equation}
		\label{eq1}
		\begin{aligned}
			\Delta v<\sqrt{\delta (v^2)}\\
		\end{aligned}
	\end{equation} 
	here $\sqrt{\delta (v^2)}$ is the velocity dispersion on the scales that turbulence is studied. In this case, intensity fluctuations in narrow channels are produced by turbulent velocities along the LOS instead of the density field. This is called the velocity caustics effect, which is a natural result of non-linear mapping from the real space to the PPV space \citep{LP00}. This effect is later absorbed into the VGT for magnetic field tracing, called the Velocity Channel Gradients (VChGs, \citealt{LY18a}).
	
	\subsubsection{The Principal Component Analysis}
	By making a synergy of the VChGs and the Principal Component Analysis (PCA), \citet{EB} produces an accurate prediction of the foreground dust polarization using the full GALFA-\ion{H}{1} and Planck data sets. \citet{2020MNRAS.496.2868L} presents a similar result. After that,
	\citet{Hu20} extends this method, i.e., VGT+PCA, to identify gravitational collapsing regions. We briefly review the recipe here.
	
	The PCA pre-processes the PPV cube to enhance the contribution from crucial components by projecting the original data set into the new orthogonal basis formed by the eigenvectors \citep{PCA}. Assuming that a PPV cube  $\rho(x, y, v)$ is a probability density function of three random variables $x$, $y$, $v$, we can obtain its covariance matrix and the eigenvalue equation for this covariance matrix from \begin{equation}
		\label{eq:13}
		\begin{aligned}
			M(v_i,v_j) \propto &\int dxdy \rho(x,y,v_i)\rho(x,y,v_j)\\
			&-\int dxdy \rho(x,y,v_i)\int dxdy\rho(x,y,v_j)
		\end{aligned}
	\end{equation}
	\begin{equation}
		\label{eq:14}
		\textbf{M}\cdot \textbf{u}=\lambda\textbf{u}
	\end{equation}
	where \textbf{M} is the co-variance matrix with $i,j=1,2,...,n_v$. $n_v$ is the number of channel in PPV cubes and $\lambda$ is the eigenvalues associated with the eigenvector $\textbf{u}$. The eigenvalues correspond to the weight of each principal component. If the eigenvalue is small, then the contribution from its corresponding principal component are also small. The eigenvectors, here, establishes a new orthogonal basis.  The projection of the PPV cube into the new orthogonal basis is operated by weighting channel $\rho(x,y,v_j)$ with the corresponding eigenvector element $u_{ij}$, in which the corresponding eigen-channel $I_i(x,y)$ is:
	\begin{equation}
		I_i(x,y)=\sum_j^{n_v}u_{ij}\cdot \rho(x,y,v_j)
	\end{equation}
	Totally $n_v$ eigen-channels in the PCA space are produced by this step. The eigen-channels are then used for gradient's calculation:
	\begin{equation}
		\label{eq:conv}
		\begin{aligned}
			\bigtriangledown_x f_i(x,y)=G_x * f_i(x,y)  \\  
			\bigtriangledown_y f_i(x,y)=G_y * f_i(x,y)  \\
			\psi_{g}^i(x,y)=\tan^{-1}[\frac{\bigtriangledown_y f_i(x,y)}{\bigtriangledown_x f_i(x,y)}]
		\end{aligned}
	\end{equation}
	where $f_i(x,y)$ represents the eigen-channel $I_i(x,y)$, $i$ = 1, 2, 3...$nv$. $\bigtriangledown_x f_i(x,y)$ and $\bigtriangledown_y f_i(x,y)$ are the $x$ and $y$ components of gradient respectively. $*$ denotes the convolution with 3 $\times$ 3 Sobel kernels $G_x$ and $G_y$\footnote{The Sobel kernels are defined as:
		$$
		G_x=\begin{pmatrix} 
			-1 & 0 & +1 \\
			-2 & 0 & +2 \\
			-1 & 0 & +1
		\end{pmatrix},
		G_y=\begin{pmatrix} 
			-1 & -2 & -1 \\
			0 & 0 & 0 \\
			+1 & +2 & +1
		\end{pmatrix}
		$$}
	. This step outputs n pixelized gradient maps $\psi_{g}^i(x,y)$, which denotes the angle of the gradients on the POS.
	
	\subsubsection{Adaptive sub-block averaging}
	
	Note that in turbulence's picture, the anisotropy of turbulent eddies is a statistic concept, in terms of the orthogonal relatively orientation of velocity gradients and the local magnetic fields. It appears only when the sampling volume is sufficient. However, each gradient in $\psi_{g}^i(x,y)$ is not required to show any correlation with the magnetic fields. To extract the anisotropy, it is necessary to obtain enough samplings over a sub-region. The corresponding algorithm was firstly proposed by \citet{YL17a}. Given a $N\times N$ pixelized gradient map, we uniformly divide the map into  
	$(N/d)\times (N/d)$ sub-blocks, where $d\times d$ is the size of each individual sub-block. The sub-block centers discretely locate at the pixel position (i$\cdot\frac{d}{2}$, j$\cdot\frac{d}{2}$), i, j=1, 3, 5, etc. The distributions of gradients' orientation within an appropriate-sized subregion appear as an accurate Gaussian profile. \citet{YL17a}, therefore, proposed the sub-block averaging
	method, i.e., taking the Gaussian fitting peak value of the
	gradient distribution in a selected sub-block to
	statistically define the mean gradient in the corresponding
	sub-block. Later instead of implementing fixed and discrete sub-blocks, \citet{Hu20} elaborated the sub-block averaging method to be adaptive. The sub-block centers were selected continuously, locating at the position (i, j), i, j=1, 2, 3, etc..The size of each sub-block is determined by the fitting errors within the 95\% confidence level. We vary the sub-block size and check its corresponding fitting errors. When the fitting error reaches its minimum value, the corresponding sub-block size is the optimal
	selection. We refer to this procedure as the adaptive sub-block (ASB)
	averaging \citep{Hu20}. 
	
	The recipe of gradient's calculation and the sub-block averaging method are applied to each eigen-channel, so that we have the eigen-gradient fields after the adaptive sub-block
	averaging $\psi_{gs}^i(x,y)$ with $i=1,2,...,n_v$. In analogy to the Stokes parameters of polarization, the pseudo Q$_g$ and U$_g$ of gradient-inferred magnetic fields are defined as:
	\begin{equation}
		\label{eq.18}
		\begin{aligned}
			& Q_g(x,y)=\sum_{i=1}^{nv} I_i(x,y)\cos(2\psi_{gs}^i(x,y))\\
			& U_g(x,y)=\sum_{i=1}^{nv} I_i(x,y)\sin(2\psi_{gs}^i(x,y))\\
			& \psi_g=\frac{1}{2}\tan^{-1}(\frac{U_g}{Q_g})
		\end{aligned}
	\end{equation}
	The pseudo polarization angle $\psi_g$ is then defined correspondingly. The pseudo polarization angle $\psi_g$ resulted from velocity gradients gives a probe of plane-of-the-sky magnetic field orientation after rotating 90$^\circ$ in the absence of self-gravity.
	
	The relative alignment between magnetic fields orientation and rotated pseudo polarization angle ($\psi_g+\pi/2$) is quantified by the \textbf{Alignment Measure} (AM, \citealt{2017ApJ...835...41G}): 
	\begin{align}
		\label{AM_measure}
		AM=2(\langle cos^{2} \theta_{r}\rangle-\frac{1}{2})
	\end{align}
	where $\theta_r$ is the angular difference in individual pixels, while $\langle ...\rangle$ denotes the average within a region of interest. AM = 1 represents the global rotated $\psi_g$ is parallel to the POS magnetic field, while AM = -1 indicates global rotated $\psi_g$ is perpendicular to the POS magnetic field. The standard error of the mean gives the uncertainty $\sigma_{AM}$; that is, the standard deviation divided by the square root of the sample size.
	
	\begin{figure*}
		\centering
		\includegraphics[width=1.0\linewidth,height=0.66\linewidth]{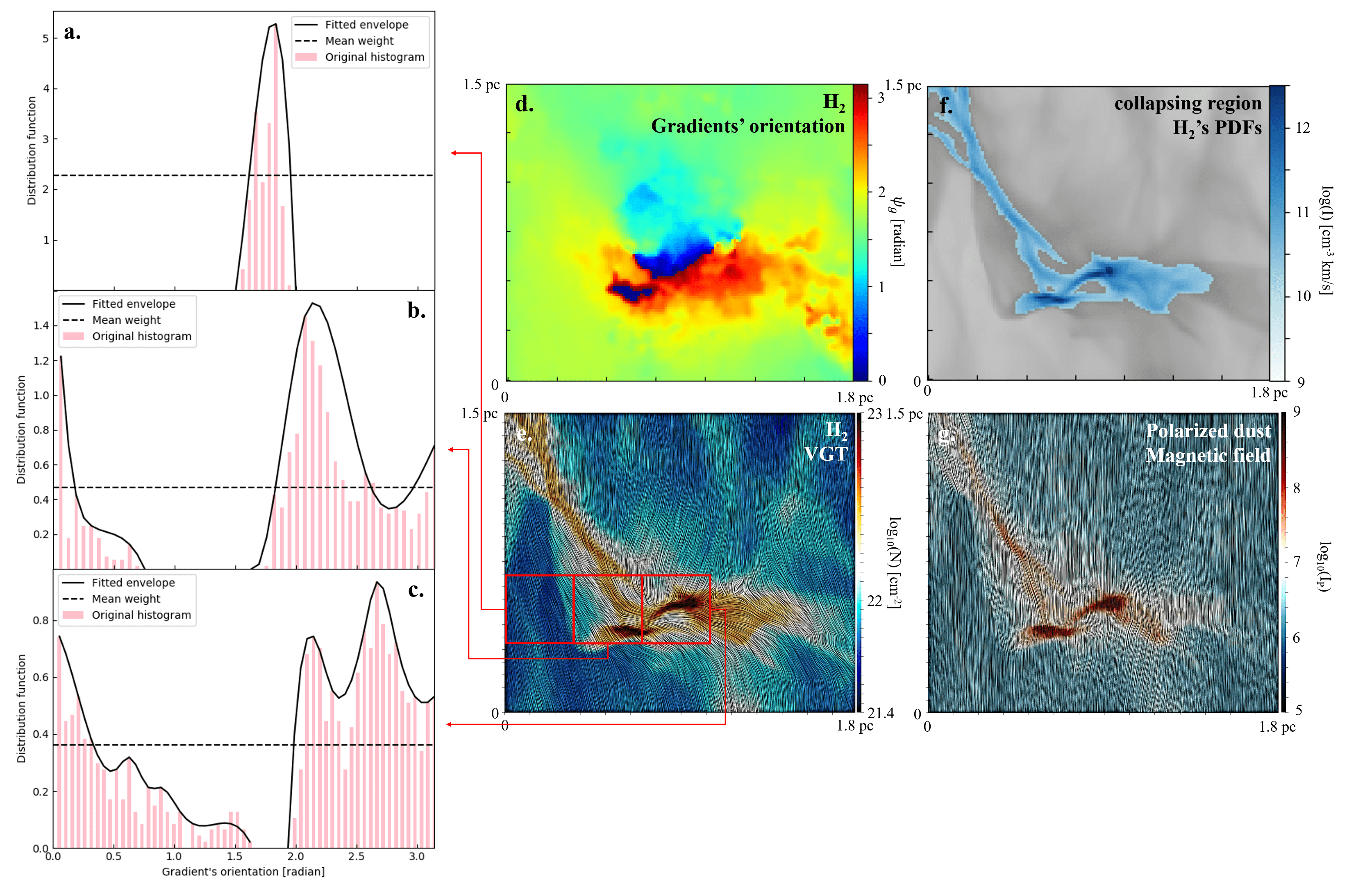}
		\caption{\label{fig:vgt} Example of how velocity gradients change orientations at the gravitational collapsing region, using simulation $\rm M_A0.4$ with t $\approx$ 0.8 Myr (see Tab.\ref{tab:sim}).
			\textbf{Panel (a) \& (b) \& (c):} histograms of velocity gradients' orientation in the diffuse region, which shows a single-peak Gaussian profile (top), in the boundary of the gravitational collapsing region, which shows a double-peak Gaussian profile (middle), in the collapsing core, which a single-peak Gaussian profile (bottom). \textbf{Panel (d) \& (e):} the map of velocity gradients' orientation in the absence of self-absorption (top). A visualization  of velocity gradient (bottom) superimposed on the projected column density map using the Line Integral Convolution (LIC). \textbf{Panel (f):} the gravitational collapsing region identified from the PDFs of $\rm H_2$ column density. \textbf{Panel (g):} the actual magnetic field morphology inferred from synthetic dust polarization. Panel (d) \& (f) are reproduced from \citet{Hu20}.
		}
	\end{figure*}
	\begin{figure*}
		\centering
		\includegraphics[width=1.0\linewidth,height=0.53\linewidth]{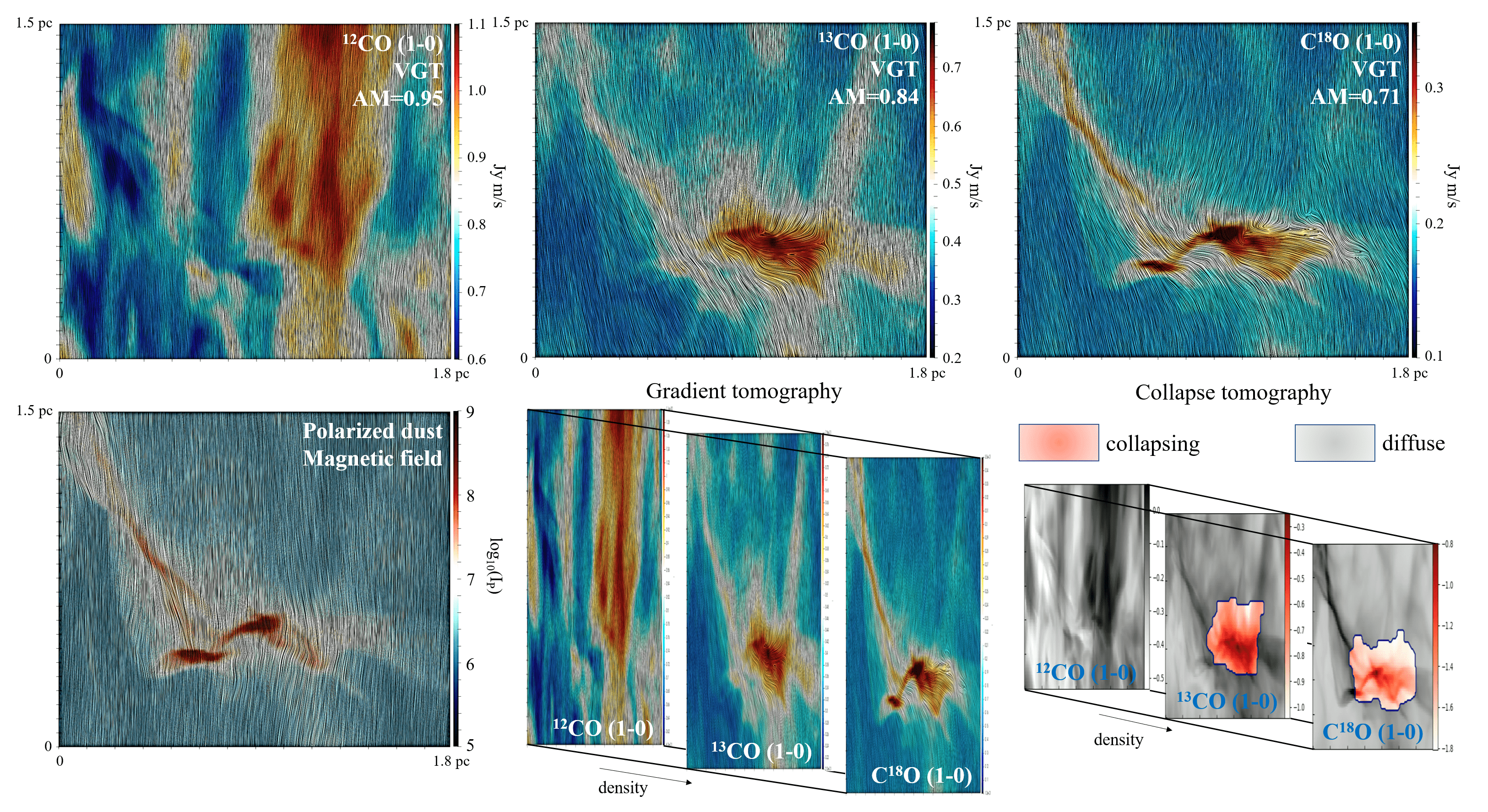}
		\caption{\label{fig:3d_vgt} Example of how velocity gradients change orientations in self-absorbing and self-gravitating region, using simulation $\rm M_A0.4$ with t $\approx$ 0.8 Myr (see Table 1). \textbf{Top:} the magnetic field morphology respectively inferred from VGT using $\rm ^{12}CO$ (left), $\rm ^{13}CO$ (middle), and $\rm C^{18}O$ (right). \textbf{Bottom:} the magnetic field morphology inferred from synthetic dust polarization (left). Middle: A three-layers gradient tomography showing the three-dimensional magnetic field orientation. Right: A three-layers collapse tomography derived from the VGT. }
	\end{figure*}
	\subsubsection{The double-peak histogram}
	Above we review the standard recipe of VGT for tracing the magnetic fields in diffuse region. However, when the self-gravity dominates over turbulence, i.e., the gravitational collapse starts, the rotated angle $\psi_g$ flips its direction by 90$^\circ$ again being perpendicular to the magnetic fields \citep{YL17b,survey,Hu20}. This change of velocity gradients can be independently used to identify the self-gravitating region. One applicable way is 
	the histogram of velocity gradients' orientation:  (i) a single peak of the histogram locates at $\psi_g$ in diffuse regions; (ii) the single peak of the histogram becomes $\psi_g+\pi/2$ in gravitationally collapsing regions; and (iii) in the transitional regions, i.e., the boundary of collapsing regions, the histogram is therefore expected to show two peak values $\psi_g$ and $\psi_g+\pi/2$, denoted as the double-peak histogram (DPH). The corresponding algorithm is presented in \citet{Hu20}. We briefly describe it here.  
	
	To implement the DPH algorithm, we define every single pixel of $\psi_g$ as the center of a sub-block and draw the histogram of velocity gradients orientation within this sub-block. Note this sub-block is different from the ASB, which is used to define the mean gradients' orientation. The second sub-block implemented in the DPH is used to extract the change of gradients. The size of the second sub-block can be different from the one for the ASB. To suppress noise and the effect from an insufficient number of bins, we plot the histogram's envelope, which is a smooth curve outlining its extremes. Any term whose histogram weight is less than the mean weight value of the envelope is masked. After masking, we work out the peak value of each consecutive profile. Once we have more than one peak value, the center of this second sub-block is labeled as the boundary of a gravitational collapsing region. By scanning all pixels in the map of velocity gradients' orientation, the boundary of self-gravitating region
	would appear as a closed contour labeled by the DPH. The region closed by the contour contains the self-gravitating gas.

	\begin{figure*}
		\centering
		\includegraphics[width=1.0\linewidth,height=0.83\linewidth]{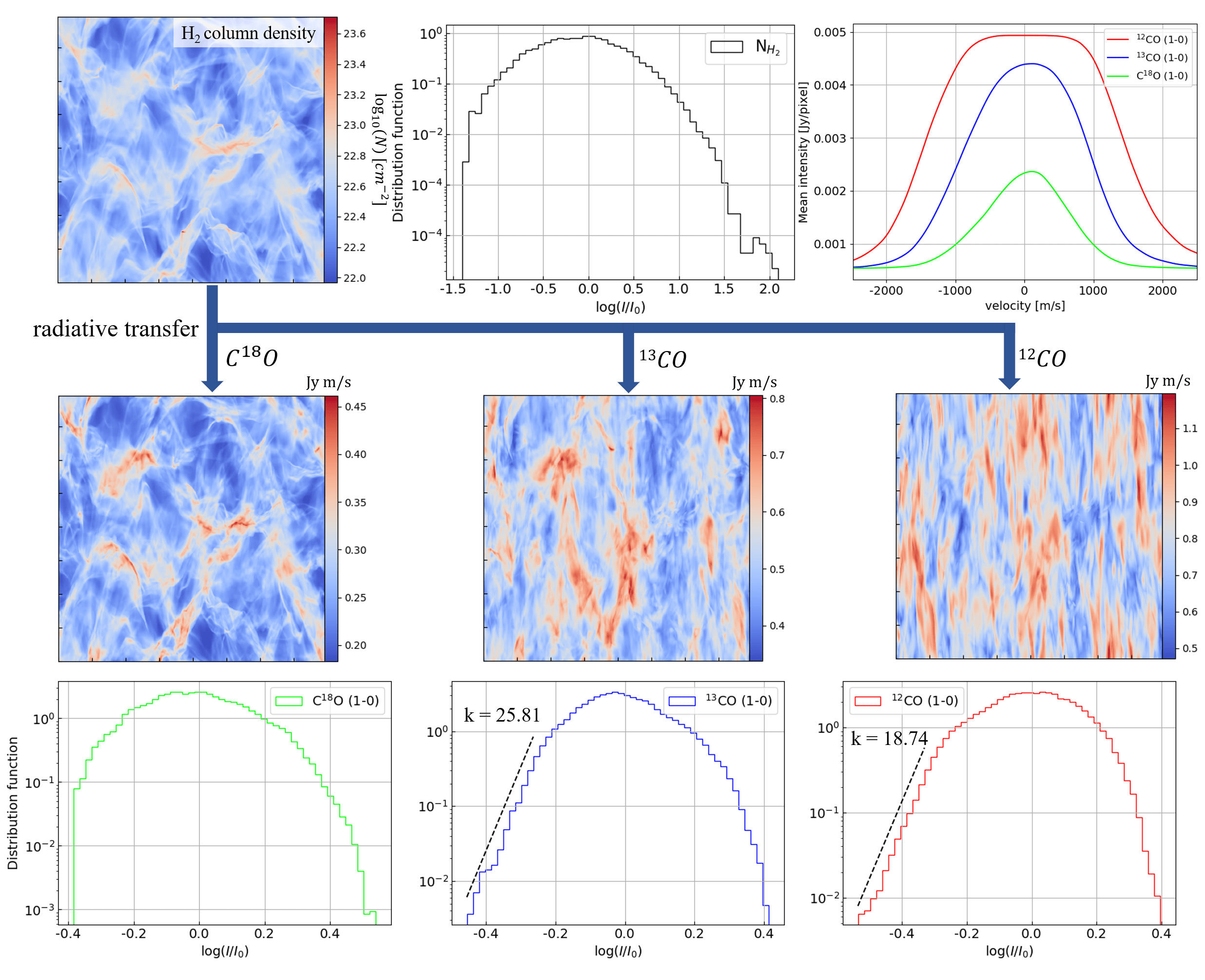}
		\caption{\label{fig:illsutration} Illustration of how the effect of radiative transfer samples the column density structures. \textbf{Top:} the actual H$_2$ column density map (left) obtained from supersonic simulation $\rm M_A0.2$ in the absence of self-gravity. We plot the corresponding PDFs (middle) and spectral lines (right). \textbf{Middle:} the integrated intensity map of three emission lines: $^{12}$CO (right), $^{13}$CO (middle), C$^{18}$O (right) in the absence of self-gravity. \textbf{Bottom:} the corresponding PDFs for each integrated intensity map. k is the fitted slope of the power-law distribution.}
	\end{figure*}
	\section{Results}
	\label{sec:results}
	\subsection{Identify gravitational collapse by the VGT}
	Fig.~\ref{fig:vgt} presents an example of how velocity gradients change their orientations at gravitational collapsing regions. We use the simulation $\rm M_A0.4$ at t $\approx$ 0.8 Myr in the absence of self-absorption. This region shows an apparent collapsing core. By using the transitional feature of the H$_2$'s PDFs, i.e., from log-normal to power-law, we identify the corresponding collapsing regions, which were previously studied in \citet{Hu20}. In the diffuse region, the resulting gradients (rotated by 90$^\circ$) are almost aligned with the magnetic field direction inferred from dust polarization. However, in the gravitational collapsing region, the gradients' orientation flips by 90$^\circ$ being perpendicular to the magnetic field. The corresponding histogram of gradients' orientation in the diffuse region exhibits only a single-peak Gaussian profile. In the collapsing region, the histogram gives a double-peak feature above the mean.
	
	Incidentally, in the upper left part, there is a filamentary structure identified as self-gravitating by the PDFs, but the gradients do not change their direction. A possible explanation could be that the density threshold given by the PDFs to distinguish high-density gas includes not only self-gravitating materials but probably also non-self-gravitating density enhancement. Also, when the volume of collapsing gas is small along the LOS, the gradients may not resolve it. 
	
	Furthermore, we consider the effect of radiative transfer in Fig.~\ref{fig:3d_vgt}. Firstly, the radiative transfer can significantly change the observed intensity structure. $\rm ^{12}CO$, which usually trace the gas density $\approx 10^{2}$ cm$^{-3}$, does not resolve the collapsing material in our simulation. Dense tracer $\rm ^{13}CO$ partially resolve the collapsing core and $\rm C^{18}O$ fully get insight into the collapsing region. The corresponding gradients of $\rm ^{12}CO$, therefore, do not show the changes of orientation but keep aligned with the magnetic field, showing AM = 0.95. Both $\rm ^{13}CO$ and $\rm C^{18}O$ exhibit the change of gradients getting lower AM values 0.84 and 0.71, respectively. These lower AM values were contributed by the self-gravity. To recover the accurate magnetic fields in the self-gravitating region, one has to identify the regions and re-rotate the velocity gradients by 90$^\circ$ again. 
	
	Here we implement the DPH algorithm to identify the collapsing regions. Note to illustrate the change of velocity gradients, we did not apply the re-rotation to Fig.~\ref{fig:3d_vgt}. The identified collapsing regions agree with the result from the H$_2$'s PDFs in Fig.~\ref{fig:vgt}. In addition, since each molecular tracer samples a different range of densities, the velocity gradients from multiple tracers tell us about the POS magnetic fields over different density ranges. For instance, in Fig.~\ref{fig:3d_vgt}, by stacking the gradient maps from $\rm ^{12}CO$, $\rm ^{13}CO$, and $\rm C^{18}O$, we can create 3D tomography information on the magnetic fields over density ranges from $10^{2}$ cm$^{-3}$ to $10^{4}$ cm$^{-3}$. The number of layers in the gradient tomography completely depends on how many molecular tracers were taken. A similar idea of collapse tomography can immigrate to the identification of collapsing regions. Here the gradient map of $\rm ^{12}CO$ does not show the feature of gravitational collapse, but the feature appears in $\rm ^{13}CO$ and $\rm C^{18}O$. The tomography can reveal the volume density range in which the collapsing occurs using multiple emission lines.

	\subsection{The intensity PDFs of CO isotopologs}
	\label{subsec: numerical results}
	
	\begin{figure*}
		\centering
		\includegraphics[width=1.0\linewidth,height=0.48\linewidth]{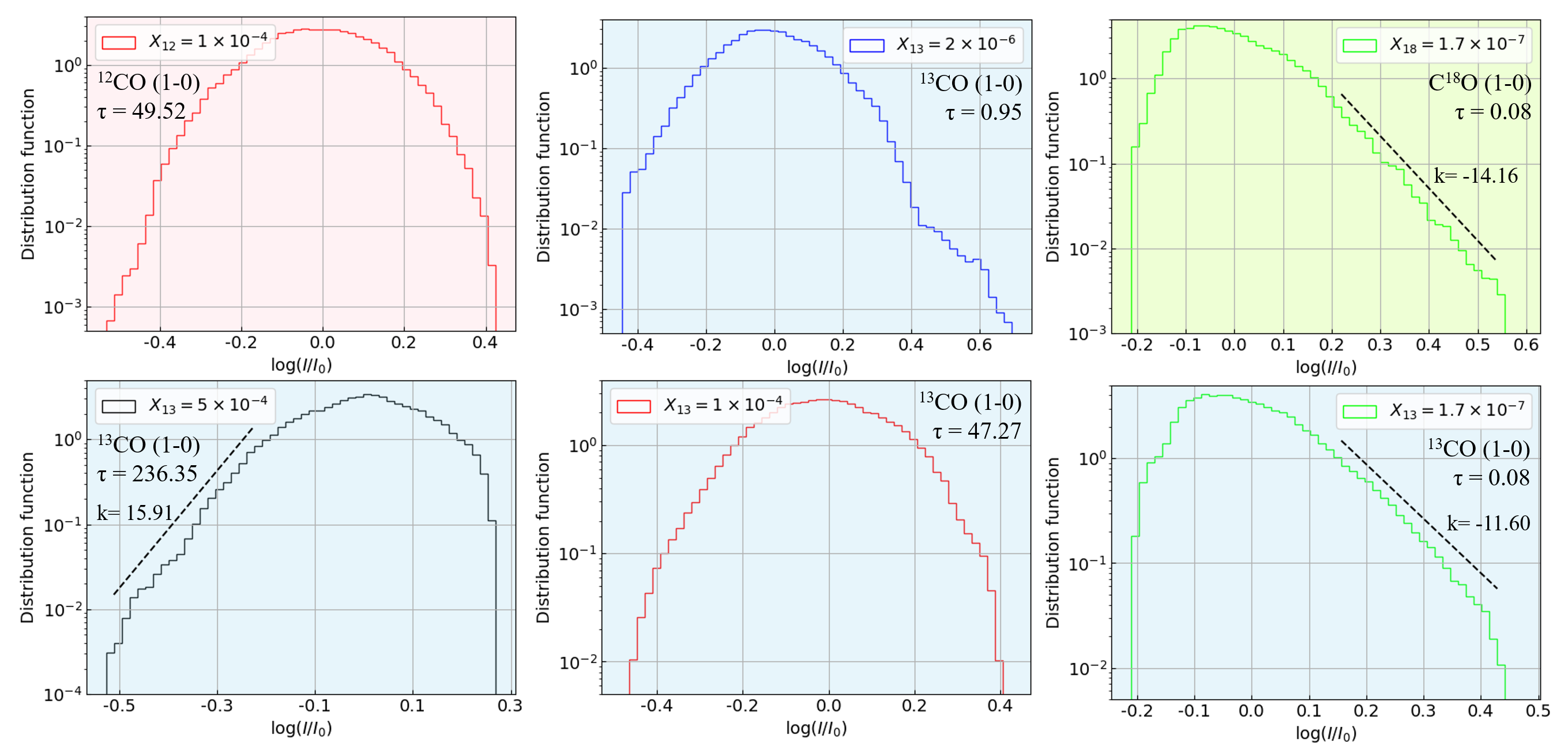}
		\caption{\label{fig:13CO} Examples of how the PDFs change their shape in the presence of self-absorption. \textbf{Top:} the PDFs of typical $^{12}$CO ($X_{12}=^{12}$CO/H$_2$ = $1\times10^{-4}$, left), $^{13}$CO ($X_{13}=^{13}$CO/H$_2$ = $2\times10^{-6}$, middle), and C$^{18}$O ($X_{18}$=C$^{18}$O/H$_2$ = $1.7\times10^{-7}$, right) media. $\tau$ gives the average optical depth along the LOS. \textbf{Bottom:} the PDFs of $^{13}$CO with various abundance: $X_{13}=^{13}$CO/H$_2$ = $5\times10^{-4}$ (left), $X_{13}=^{13}$CO/H$_2$ = $1\times10^{-4}$ (middle), $X_{13}=1.7\times10^{-7}$ (right).
		}
	\end{figure*}
	
	\begin{figure}
		\centering
		\includegraphics[width=1.0\linewidth,height=0.73\linewidth]{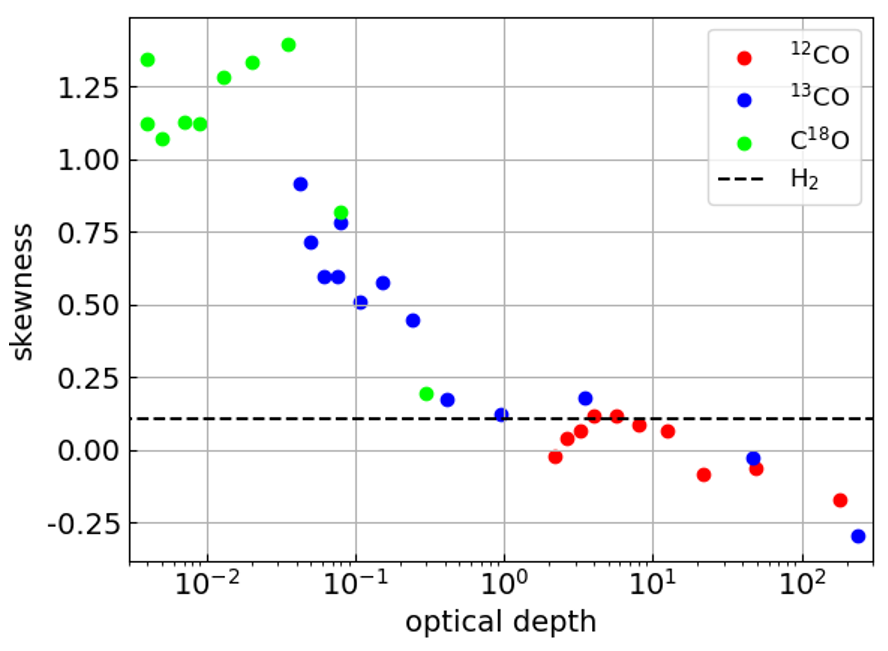}
		\caption{\label{fig:sk} The correlation of skewness and optical depth $\tau$. We use the synthetic maps of $^{12}$CO (red), $^{13}$CO (blue), C$^{18}$O (lime) generated by super-sonic simulations $\rm M_S\approx 6$. The dashed line denotes the mean skewness of H$_2$'s PDFs.
		}
	\end{figure}
	
	Fig.~\ref{fig:illsutration} shows an example of how the intensity PDFs change their shape in the presence of self-absorption. We consider the supersonic simulation $\rm M_A0.2$ here. As expected, $^{12}$CO, $^{13}$CO, and C$^{18}$O typically trace gas with H$_2$ number density between $10^2$ and $10^4$ cm$^{-3}$, i.e., $10^2$ cm$^{-3}$ for $^{12}$CO, $10^3$ cm$^{-3}$ for $^{13}$CO, and $10^4$ cm$^{-3}$ for C$^{18}$O. The inefficiency of a saturated $^{12}$CO emission line cannot trace the density range above the saturation threshold, while C$^{18}$O traces only materials with high-density.
	
	We plot the normalized PDFs for H$_2$ and CO isotopologs intensity maps in Fig.~\ref{fig:illsutration}. The PDFs' bin size is selected as 50, which is accurately sufficient for a high-resolution simulation. Note here we focus on the shape of the PDFs rather than its dispersion.  We did not constrain the PDFs to an identical intensity range. To distinguish column density PDFs and intensity PDFs, we denote s$_c=ln(N/N_0)$ for H$_2$ column density and s$_i=ln(I/I_0)$ for CO isotopologs' intensity, where $I_0$ and $N_0$ are the mean intensity and mean density respectively. In the case of H$_2$ column density, the PDFs exhibit an expected log-normal distribution, and its width is controlled by $\rm M_S$. Importantly, the log-normal PDFs are insensitive to shocks. This property boosts the synergy of PDFs and VGT in terms of identifying the gravitational collapse and distinguish shocks. For instance, the shock density jump condition induces a dominated density gradient perpendicular to shock fronts. Incidentally, the magnetic fields suppress the compression in its perpendicular direction so that the shock fronts are predominately perpendicular to the field lines \citep{2019ApJ...878..157X}. As a result, the density gradient also flips its direction from orthogonal to align with the magnetic fields, which may confuse the density gradient change induced by self-gravity. \citet{IGs} proposed one solution to distinguish shocks by comparing the velocity gradients and the density gradients since the velocity field is less sensitive to shocks. Here we see the PDFs provide one more solution. When the density gradient's direction gets changed, but the corresponding density PDFs is log-normal, it indicates shocks' presence. Similarly, when the change of direction and power-law PDFs appears, it corresponds to the case of gravitational collapse.

	The PDFs become complicated when considering the effect of radiative transfer. For the supersonic simulation $\rm M_A0.2$, the intensity PDF obtained from $^{12}$CO is not a log-normal distribution, but a power-law distribution towards the low-intensity range. We perform a least-squares fitting to the power-law parts. The fitting error is given by the margin error within a 95\% confidential level. The fitting slope of the power-law distribution is $k=18.74\pm1.05$. The intensity PDF of $^{13}$CO also shows a power-tail in the low-intensity range with a steeper slope  $k=25.81\pm1.03$. While for the intensity PDF of C$^{18}$O, it is more close to a single log-normal distribution, similar to the case of H$_2$. Above we consider only the effect of radiative transfer in the absence of self-gravity. 
	
	As discussed above, the shape of the PDFs gets changes under radiative transfer. Here, instead of using different emission lines, we vary the abundance of $^{13}$CO in simulation $\rm M_A0.4$, i.e., the ratio $^{13}$CO/H$_2$, which changes the optical depth. We use $^{13}$CO/H$_2$ = $1\times10^{-4}$ ( $\tau = 47.27$) which is the typical value of $^{12}$CO abundance, typical $^{13}$CO/H$_2$ = $2\times10^{-6}$, $^{13}$CO/H$_2$ = $5\times10^{-4}$ ($\tau = 236.35$), and $^{13}$CO/H$_2$ = $1.7\times10^{-7}$ ($\tau = 0.08$) which is the typical value of C$^{18}$O abundance (see \S~\ref{Sec.data}). 
	
	In Fig.~\ref{fig:13CO}, for the typical $^{12}$CO ($^{12}$CO/H$_2$ = $1\times10^{-4}$) and $^{13}$CO ($^{13}$CO/H$_2$ = $2\times10^{-6}$) media , we see that their PDF is similar to a single log-normal distribution. The PDF of C$^{18}$O, however, exhibits a power-law tail at high-intensity range with a slope $k=-14.16$. To investigate the origin of this power-law tail, then we use only $^{13}$CO and change its abundance to $^{13}$CO/H$_2$ = $1.7\times10^{-7}$. The lower abundance shapes the PDF of $^{13}$CO to a hybrid of log-normal distribution in the low-intensity part and power-law distribution in the high-intensity part, which is indeed the case of C$^{18}$O. The slope of the power-law part, however, gets shallower ($k=-11.60$). It implies that the difference in CO isotopologs also contribute to the PDFs, but not significant. Also, we adopt the abundance $^{13}$CO/H$_2$ = $1\times10^{-4}$, the resulting PDF of $^{13}$CO has its shape similar to the one of $^{12}$CO. When the abundance is increased further to $^{13}$CO/H$_2$ = $5\times10^{-4}$, a low-intensity power-law tail appears in the PDF.
	
	In this test, we use a single molecular tracer but vary its abundance, i.e., optical depth, which excludes molecules' different chemical properties. We still see the change of intensity PDFs using a single $^{13}$CO tracer. Therefore, it confirms that the appearance of low-intensity tail and high-intensity tail is caused by the different sampling powers of molecular tracers. 
	
	\subsection{Skewness of the PDFs}
	To quantify the PDFs' shape, we calculate the skewness of H$_2$ and CO isotopologs' intensity. The skewness $\gamma$ of a data sample is statistically defined as:
	\begin{equation}
		\gamma=\frac{1}{N}\sum^{N}_{i=1}(\frac{s_i-\langle s\rangle}{\sigma_s})^3
	\end{equation}
	where N is the sample size, $\langle s\rangle$ is the mean value, and $\sigma_s$ is the standard deviation. Skewness can be quantified as a representation of the extent to which a given distribution varies from a normal distribution. Negative skewness refers to a longer or fatter tail on the left side of the PDF, while positive skewness refers to a longer or fatter tail on the right. The uncertainty of skewness is related to the sample size as $\pm\sqrt{\frac{24N(N-1)}{(N-2)(N+1)(N+3)}}$, which is negligible in our study due to large data sampling.
	\begin{figure}
		\centering
		\includegraphics[width=1.0\linewidth,height=2.66\linewidth]{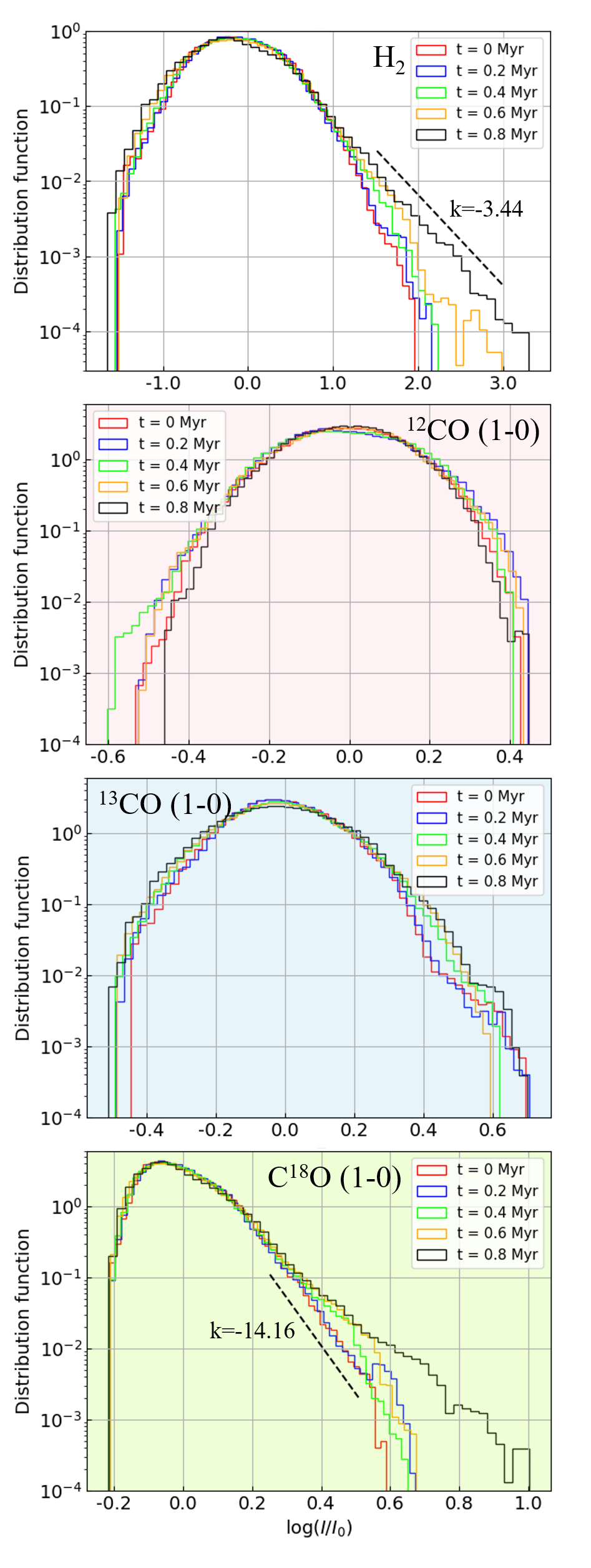}
		\caption{\label{fig:pdf_grav} Examples of how the PDFs change their shape in the presence of radiative transfer and self-gravity. The simulations $\rm M_A$0.4 are used here. k is the slope of the power-law distribution.
		}
	\end{figure}
	
	In Fig.~\ref{fig:sk}, we plot the correlation of skewness and $\tau$.
	We calculate the mean skewness of H$_2$'s PDFs over ten MHD simulations, getting the value is around 0.106. We use this skewness value as a threshold above or below which the PDFs' shape starts evolving. In particular, We find skewness is generally decreasing with the increment of $\tau$, regardless of the CO isotopologs. In the range of $\tau<10$ the skewness is positive, while when $\tau>10$, the skewness gradually deviates towards negative, which indicates the low-intensity power-law tail starts appearing. In the situations of $\tau<0.3$, the skewness becomes larger than the threshold value 0.106, which means the high-intensity power-law tail starts appearing. Consequently, we expect that the optically thick $^{12}$CO always shows intensity PDFs with a low-intensity power-law tail for observational data. As for the optically thin media, the intensity PDFs are in log-normal shape when approximately $0.3<\tau<10$, while the intensity PDFs exhibit high-intensity power-law tails in the case $\tau<0.3$.
	
	\subsection{Intensity PDFs of self-gravitating and self-absorbing media}
	Self-gravity plays a vital role in the shape of column density PDFs. In the case of self-gravitating media, the column density PDFs are expected to be a power-law tail in the high-density range. In the previous section, we find the high-density tracers C$^{18}$O with small optical depth also produce the same high-intensity tail. We, therefore, explore how self-gravity contribute to the intensity PDFs of self-absorbing media.
	
	We present the resulting PDFs of self-gravitating CO isotopologs in Fig.~\ref{fig:pdf_grav}. The column density PDF of H$_2$ is initially log-normal. With the increment of self-gravity, which is proportional to running time t, the PDF exhibits a power-law tail in high-density range. At t = 0.8 Myr, we have a slope $k=-3.44$ for the power-law tail. The PDFs of $^{12}$CO and $^{13}$CO, however, always keep an approximately log-normal shape, regardless of the self-gravity. The width of the PDFs keeps a constant value $\approx 1$. As these two species samples more diffuse regions, we expect the sonic Mach number controls the width. In addition, the independence of self-gravity also appears in C$^{18}$O's PDFs. Its PDFs initially show a power-tail in high-density range, which is similar to the case of H$_2$. This power-tail is fixed with a slope $k=-14.16$ until t = 0.2 My and then it gets shallower at t = 0.8 Myr. It indicates the sufficiently strong self-gravity can still shape the PDFs of C$^{18}$O which usually samples dense gas. However, it is difficult to distinguish the self-gravity induced power-law tail and the self-absorption induced power-law tail.  
	\begin{figure}
		\centering
		\includegraphics[width=1.0\linewidth,height=0.75\linewidth]{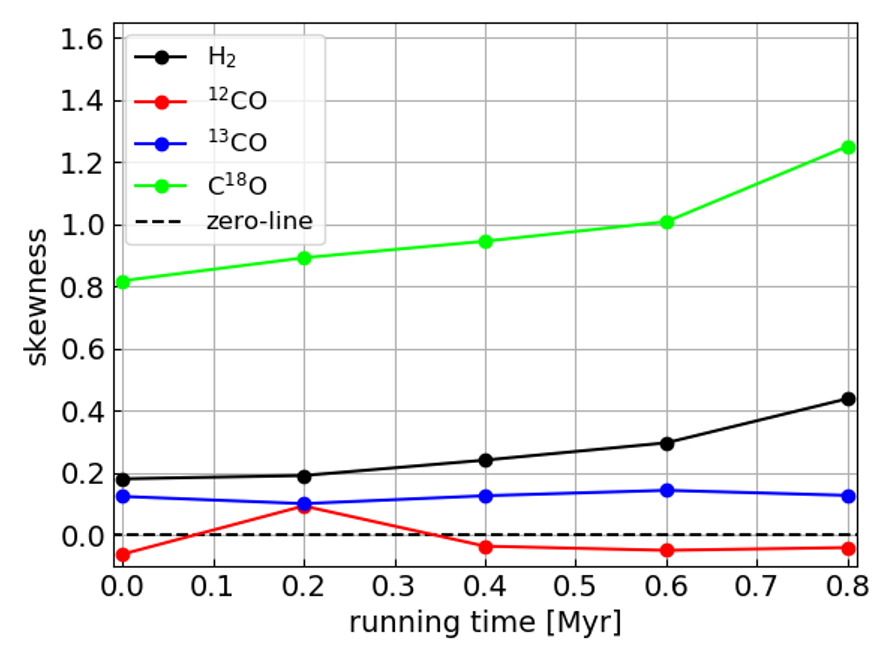}
		\caption{\label{fig:sk_grav} The skewness calculated from simulations $\rm M_A0.4$ at different snapshots in the presence of self-gravity.}
	\end{figure}
	
	In Fig.~\ref{fig:sk_grav}, we calculate the skewness of each PDF used in Fig.~\ref{fig:pdf_grav}. We can see H $ _2$'s PDFs' skewness is positive and positively proportional to the running time, i.e., the strength of self-gravity. However, in the cases of $^{12}$CO and $^{13}$CO, the skewness is stably staying at 0.15 around for $^{13}$CO and -0.05 around for $^{12}$CO. The skewness of $^{12}$CO and $^{13}$CO is insensitive to the self-gravity. However, the skewness of C$^{18}$O gets significantly approximately 3.5 times higher value than the H $ _2$ cases. This large skewness implies a steep power-tail in high-intensity range, as shown in Fig.~\ref{fig:pdf_grav}. Therefore, using intensity PDFs, there is difficulty in distinguishing the power-law produced by self-gravity or radiative transfer.

	\section{Observational results}
	\label{sec:obs}
	\subsection{The PDFs of $\rm ^{12}CO$, $\rm ^{13}CO$, and H$_2$ }
	To demonstrate the change of density PDFs in the presence of self-absorption in observation, we utilize two emissions lines $\rm ^{12}CO$ (2-1) and $\rm ^{13}CO$ (2-1) of molecular cloud NGC 1333. NGC 1333 is one of the most active star-forming clouds in the solar vicinity with an average H$_2$ volume density $\rm n(H_2)\approx 1750 cm^{-3}$ \citep{1996A&A...306..935W}. It locates at a distance of $\simeq 235$ pc as a section of the Perseus molecular cloud. The emissions lines are mapped from the Arizona Radio Observatory CO Mapping Survey with the Heinrich Hertz Submillimeter Telescope \citep{2014ApJS..214....7B}. At the same time, the H$_2$ column density data is obtained from the Herschel Gould Belt Survey \citep{2010AA...518L.102A}. 
	
	The angular resolution of the emission liens is 38$''$ (0.04 pc), and velocity resolution is 0.3 km $\rm s^{-1}$ with a sensitivity of 0.15 K RMS noise per pixel in one spectral channel \citep{2014ApJS..214....7B}. The radial velocity of the bulk of the emission ranges from about -4 to +16 km $\rm s^{-1}$ for $\rm ^{12}CO$ (2-1) and from -1 to +11 km $\rm s^{-1}$ for $\rm ^{13}CO$ (2-1) \citep{2014ApJS..214....7B}. We select the emissions within these ranges for our analysis.
	
	\begin{figure}
		\centering
		\includegraphics[width=1.0\linewidth,height=2.15\linewidth]{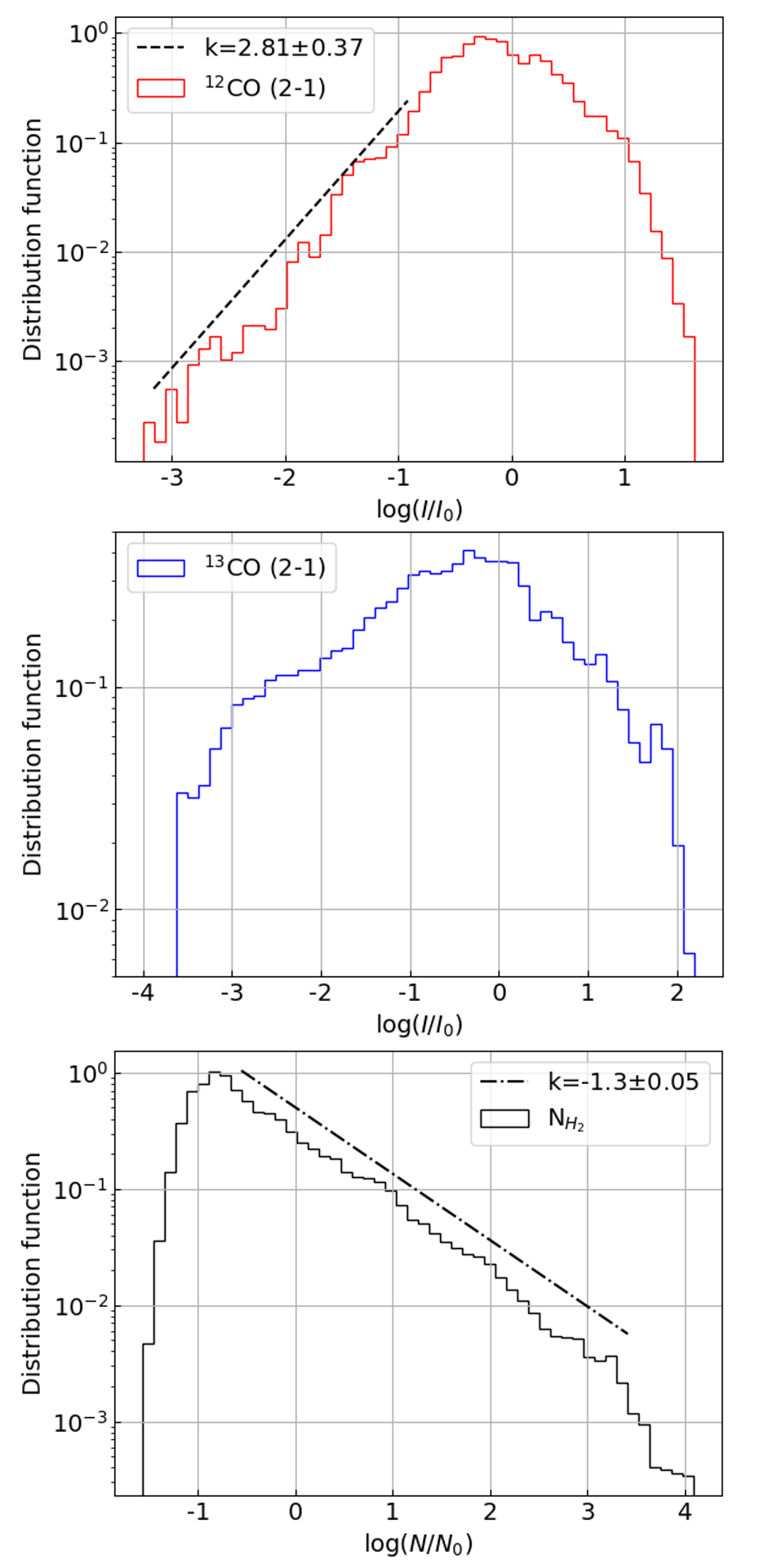}
		\caption{\label{fig:ngc1333_pdf}Observational examples of how the intensity PDFs change their shape for molecular cloud NGC 1333: $\rm ^{12}CO$ (top), $\rm ^{13}CO$ (middle), and H$_2$ (bottom). The dashed black line is the best fitting for the power-law part of PDFs within 95\% confidential level and $k$ denotes its slope.}
	\end{figure}
	
	\begin{figure}
		\centering
		\includegraphics[width=1.0\linewidth,height=1.76\linewidth]{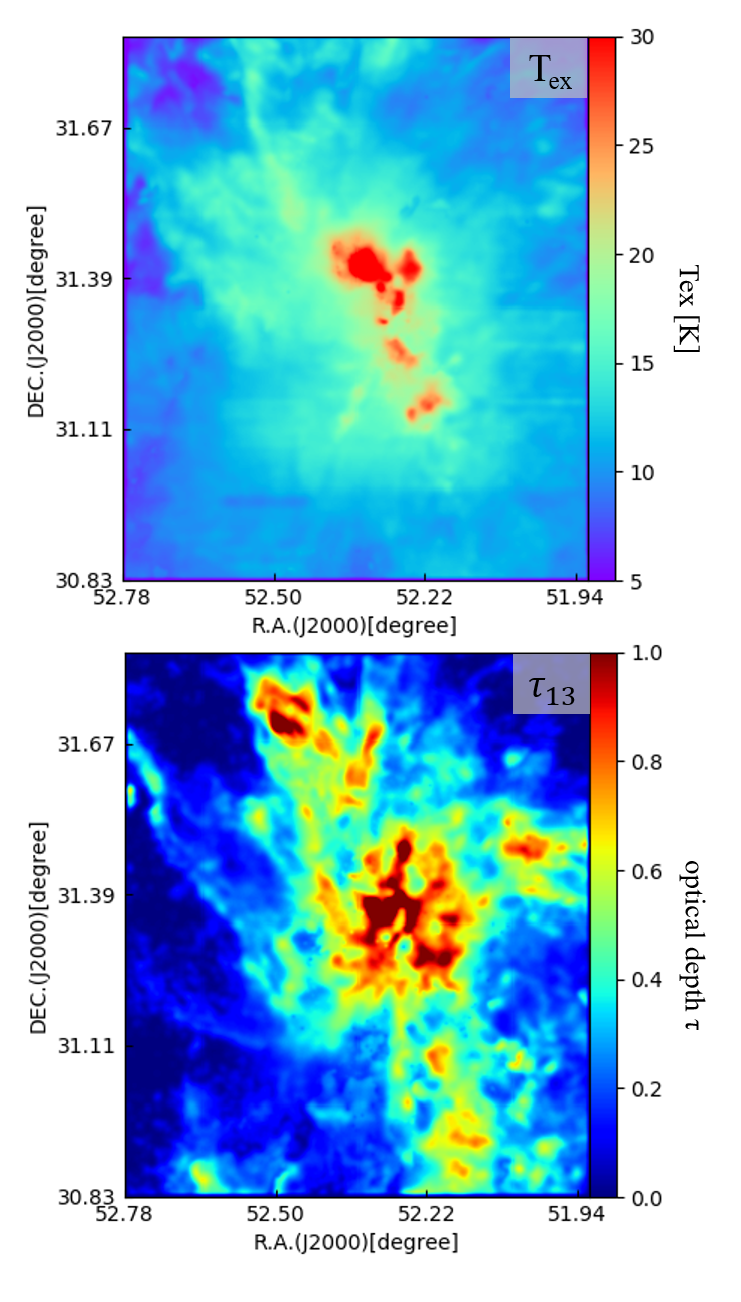}
		\caption{\label{fig:ngc1333_tau} Maps of excitation temperature $\rm T_{ex}$ (top) and the optical depth $\tau_{13}$ (bottom) of molecular cloud NGC 1333.}
	\end{figure}

	\begin{figure}
		\centering
		\includegraphics[width=1.0\linewidth,height=0.76\linewidth]{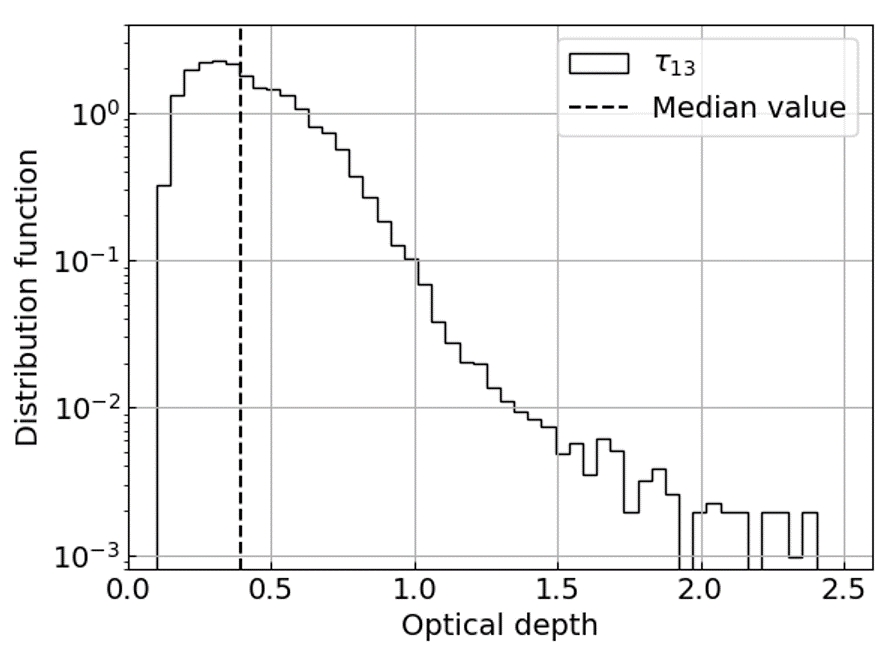}
		\caption{\label{fig:hist_tau} Histogram of the derived optical depth $\tau_{13}$ for molecular cloud NGC 1333.}
	\end{figure}
	
	In Fig.~\ref{fig:ngc1333_pdf}, we plot three intensity PDFs of each molecular tracer $^{12}$CO (2-1), $^{13}$CO (2-1), and H$_2$ column density respectively. Pixels where the brightness temperature is less than 0.45 K, which is about 3 times the RMS noise level, are blanked out. For the intensity PDF of $^{12}$CO (2-1), we see it appears as a power-law at low-intensity regions with the slope $k=2.81\pm0.37$. The slope is shallower than our numerical results. It likely other factors also contribute to this power-law tail, such as the effects of line-of-sight contamination \citep{2015A&A...575A..79S,2019MNRAS.484.3604L} and stellar feedback via ionization \citep{2014A&A...564A.106T}.
	
	However, the intensity PDF of $^{13}$CO (2-1) is more similar to a log-normal distribution. As for H$_2$ column density, its PDF becomes completely power-law at the high-intensity region. The slop of this power-law part is $k=-1.3\pm0.05$.  Similar to the skewness of numerical data, the skewness of $^{12}$CO is negative $\approx-0.15$, while for $^{13}$CO is approximately 0. 
	
	This power-law transition seen in the H$_2$'s PDF is believed to indicate gravitational collapse. The characteristic slope of that power-law changes in roughly the cloud means free fall time from steep, $k\approx-3$, to shallow values, $k\approx -1.5$ \citep{2014ApJ...781...91G,2017ApJ...834L...1B,2018MNRAS.477.5139G} . The broad range of $k$ is regulated by both of the magnetic fields and the efficiency of feedback \citep{2015MNRAS.450.4035F,2015ApJ...808...48B,2017ApJ...840...48P}. In our case, the critical threshold for the power-law begins to form is rough $\log(N/N_0)\approx-0.8$ and $k=-1.3\pm0.05$, which means the majority of the NGC 1333 cloud has collapsed into isothermal cores.
	
	However, this power-law feature is observed in neither $^{12}$CO nor $^{13}$CO. The negative power-law transition in the low-density range of $^{12}$CO's PDF and the log-normal PDFs of $^{13}$CO provide little information about the gravitational collapse. According to our numerical analysis, these features are resulting from the effect of radiative transfer. To test this point, we further calculate the optical depth $\tau_{12}$ and $\tau_{13}$ for $^{12}$CO (2-1) and $^{13}$CO (2-1), respectively.
	
	\subsection{Optical depth of $^{13}$CO}
	To derive the optical depth of $^{13}$CO, we assume that $^{12}$CO is optically thick, $^{12}$CO, and $^{13}$CO trace the
	same component, and these lines reach local thermal
	equilibrium. Thus, the temperature of $^{12}$CO (2-1) can be
	treated as the excitation temperature, $\rm T_{ex}$, of $^{13}$CO (2-1) for deriving the optical depths. Then, we obtain their peak intensities by applying Gaussian fitting to spectra of the $^{12}$CO and $^{13}$CO cube data for pixels with their intensity $\ge0.45$ K, which is about three times the RMS noise level.
	
	Fig.~\ref{fig:ngc1333_tau} shows the excitation temperature map derived from the following equation with the previous assumptions and the beam filling factor of one\citep{2010ApJ...721..686P,2015ApJ...805...58K,1976ApJS...30..247U}:
	\begin{equation}
		\begin{aligned}
			J_\nu(T)&=\frac{h\nu/k_B}{exp(h\nu/(k_B T))-1}\\
			T_{ex}&=\frac{h\nu_{12}}{k_B}[\log(1+\frac{h\nu_{12}/k_B}{T_{mb}^{12}+J_\nu(T_{bg})})]^{-1}\\
			&=11.06[\log(1+\frac{11.06}{T_{mb}^{12}+0.194})]^{-1}
		\end{aligned}
	\end{equation}
	where $h$ is the Planck constant, $\nu_{12}$ is the emission frequency of $^{12}$CO (2-1), $k_B$ is the Boltzmann constant,  $T_{mb}^{12}$ is the peak intensity of $^{12}$CO (2-1) in units of K from the above fitting, $J_\nu(T)$ is the effective radiation temperature, and $T_{bg}$ = 2.725 K is the temperature of cosmic microwave background radiation. The optical depths $\tau_{13}$ of the $^{13}$CO (2-1) is derived from the following equations \citep{1994ApJ...429..694L,Kawamura1998}:
	\begin{equation}
		\tau_{13}=-\log[1-\frac{T_{mb}^{13}/\phi_{13}}{10.58([\exp(10.58/T_{ex})-1]^{-1}-0.02)}]
	\end{equation}
	where $T_{mb}^{13}$ is the peak intensities of $^{13}$CO (2-1) in units of K and $\phi_{13}$ is the beam filling factors. The beam filling factors can be expressed as $\phi_{13}=\theta^2_{source}/(\theta^2_{source}+\theta^2_{beam})$, where $\theta^2_{source}$ and $\theta^2_{beam}$ are the source size and the effective beam size, respectively. Since the effective beam sizes we used here are 38$''$ (corresponding to 0.04 pc), which is much smaller than the typical core size, we simply adopt $\phi_{13}=1$.
	
	We derived a median excitation temperature $T_{ex}\approx13.04$ K.
	Fig.~\ref{fig:ngc1333_tau} presents the optical depth maps of $^{13}$CO (2-1). The optical depth is more than 1 at the
	center of NGC 1333 and drops to less than one at the outer edge.
	As shown in Fig.~\ref{fig:hist_tau}, $\tau_{13}$ is in the range of $0.18\le\tau_{13}\le2.46$ with a median value $\tau_{13}\approx0.45$. By adopting the intrinsic ratio of abundances  $\rm f_{12/13} = 50\pm10$ for the interstellar gas in the solar neighborhood \citep{2005ApJ...634.1126M}, we obtain $9.0\pm1.8\le\tau_{12}=f_{12/13}\tau_{13}\le123.0\pm24.6$. This high optical depth indicates that $^{12}$CO (2-1) is probably self-absorbed, as suggested by \citet{2014ApJS..214....7B}. The PDF optically thick media $^{12}$CO presents a power-law tail at low-intensity range and the PDF of optically thin media $^{13}$CO gives is more close to a log-normal distribution. 
	
	\begin{figure}
		\centering
		\includegraphics[width=1.0\linewidth,height=1.75\linewidth]{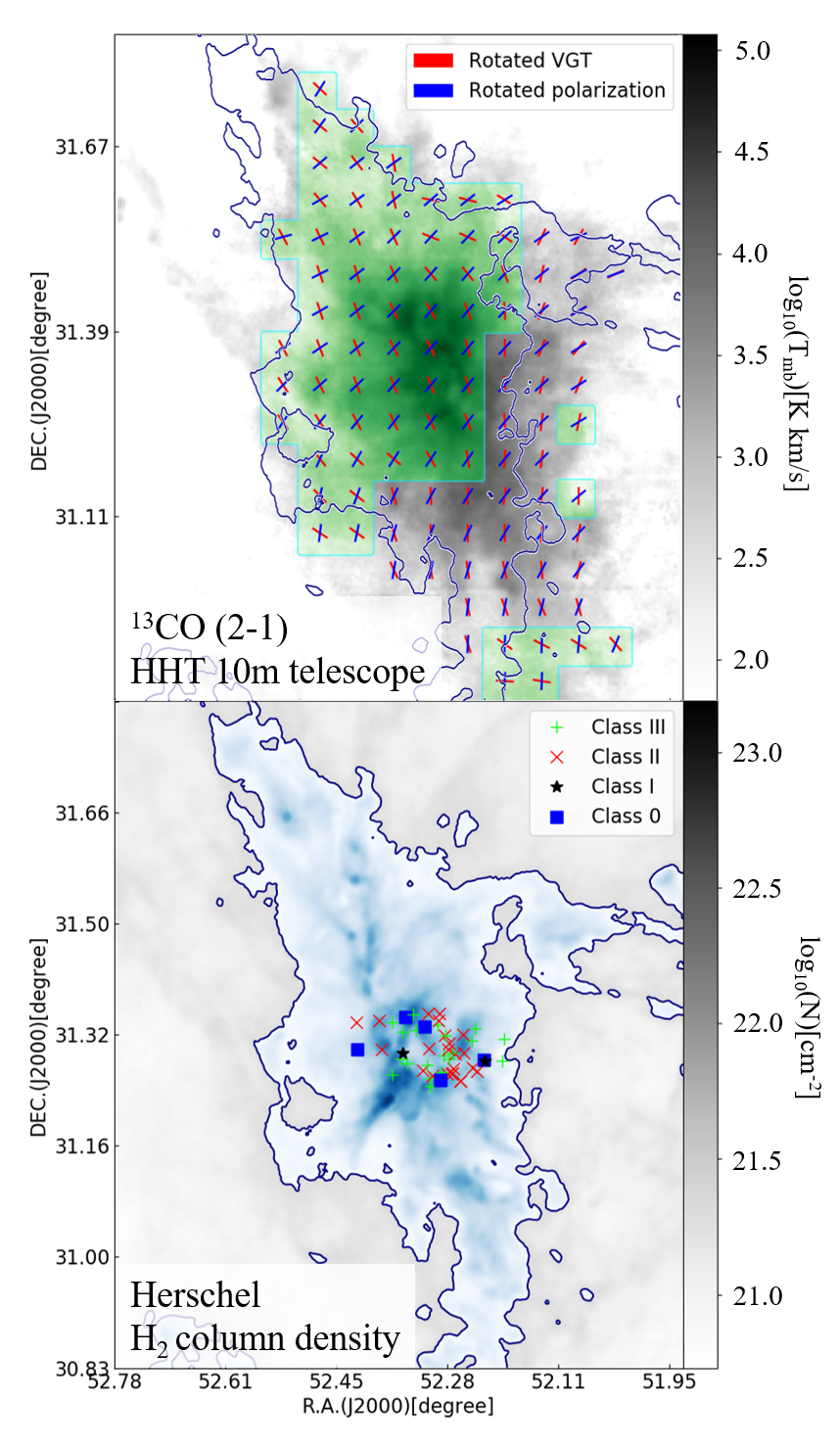}
		\caption{\label{fig:ngc1333_vgt}\textbf{Top:} Observational results of the magnetic field derived from VGT (red segment) and Planck polarization (blue-segment) in molecular cloud NGC 1333. Green areas (sharing colorbar with gray background) are self-gravitating regions (e.g., AM $<$ 0) identified by VGT. Dark blue contour outlines the self-gravitating regions identified by PDFs. \textbf{Bottom:} The H$_2$ column density map obtained from the Herschel Gould Belt Survey \citep{2010AA...518L.102A}. Blue areas (sharing colorbar with gray background) are self-gravitating regions identified from column density PDFs. The YSOs are identified by \citet{2010AJ....140..266W}.}
	\end{figure}
	
	\subsection{Application of VGT in NGC 1333}
	Here we present the results of VGT using emission lines $\rm ^{13}CO$ (2-1). Pixels where the integrated brightness temperature is less than 6 K (3$\sigma$ level) are blanked out. We also make a comparison with the Planck 353GHz polarized dust signal data from the Planck 3rd Public Data Release (DR3) 2018 of High-Frequency Instrument \citep{2018arXiv180706207P}. The Planck observations provide Stokes parameter maps I, Q, and U, so the POS magnetic field orientation angle $\theta$ in IAU convention can be derived from the Stokes parameters: $\theta=\frac{1}{2}\arctan(-U, Q)-\frac{\pi}{2}$. In this work, we smooth the Q, U maps through a Gaussian filter with the FWHM $\approx$ 10 arcsec. The corresponding study in \citet{2020A&A...641A..12P} shows the uncertainty of polarization fraction $\sigma_p$ is less than 0.002\% in the Galactic plane $|b|<$30$^\circ$ (see Fig.~2 in \citealt{2020A&A...641A..12P} ). The minimum value polarization percentage of $p$ in NGC 1333 is around 0.05\%. The signal-to-noise ratio $p/\sigma_p>3$, therefore overs most of the cloud.  
	
	In Fig.~\ref{fig:ngc1333_vgt}, we observed the resulting rotated velocity gradients are nearly perpendicular to the magnetic fields inferred from the Planck polarization. This orthogonal relative orientation indicates the majority of the cloud is under gravitational collapse, which agrees with the result of the H$_2$'s PDF (see Fig.~\ref{fig:ngc1333_pdf}). Here we did not apply the DPH algorithm to identify the boundary of collapsing regions. The DPH implementation requires that the observational data covers sufficient diffuse regions so that the gradients give apparent transition. As the NGC 1333 data here covers only the high-density self-gravitating regions. It does not meet the requirement of DPH. In addition, this cloud has been studied by \citet{survey}, which used only one central thin channel for analysis, showing AM = -0.71. Here, we introduce the PCA to pre-process the spectroscopic data and construct the pseudo-Stokes parameters (see Eq.~\ref{eq.18}). These producers consider all thin velocity channels, including both central and wing channels. Comparing with \citet{survey}, these additional procedures yield AM $\approx$ -0.45. The higher AM value is contributed by mainly three factors. One is the diffuse gas in wing channels, which give positive AM. Also, the PCA which extracts the most important velocity components suppresses the noise and increases the alignment \citep{PCA}.  The pseudo-Stokes parameters here sum the gradients in a way similar to the polarization measurement, instead of pure vector summation. The pseudo-Stokes parameters improve the alignment between the gradients and the magnetic fields inferred from the dust polarization \citep{EB,2020MNRAS.496.2868L}. Therefore, we expect a better alignment between the gradients and the magnetic fields with these additional procedures. Nevertheless, the negative AM still suggests that the NGC 1333 cloud is self-gravitating.
	
	The above conclusion about NGC 1333 can be also demonstrated by the H$_2$'s PDF (see Fig.~\ref{fig:ngc1333_pdf}). The entire PDF exhibit an apparent power-law distribution, which means the self-gravitating gas occupies a large fraction of NGC 1333. By adopting the transition threshold $\log(N_{st}/N_0)\approx0.5$, we identify the self-gravitating regions in Fig.~\ref{fig:ngc1333_vgt}. As a comparison, we use green color to outline the self-gravitating regions (AM$<$0) identified by VGT. We find in north and central areas, VGT and PDFs suggest the presence of gravitational collapse. We also plot the Young Stellar Objects (YSOs) identified by \citet{2010AJ....140..266W}. The highly clustered YSOs in the center also suggest NGC 1333 is actively forming stars. However, VGT reveals a quiescent south tail while PDFs reject this result. The south tail is likely only a high-density enhancement. However, the high-density enhancement is overwhelmed in the PDFs' power-law part. Therefore, the PDFs' density threshold statistically includes both self-gravitating material and non-self-gravitating density enhancement. It is also possible the collapse happens at a higher density range ($n\ge10^3$ $\rm cm^{-3}$) so that $\rm ^{13}CO$ does not resolve it.


	\section{Discussion} \label{sec.dis}
	\subsection{Synergy of VGT and PDFs}
	The probability density functions  (PDFs)  of the column mass density are widely used to study turbulence properties and self-gravity in ISM. It was shown that the column density PDFs are log-normal in diffuse isothermal clouds \citep{1995ApJ...441..702V,2000ApJ...535..869K,2008ApJ...680.1083R,2011ApJ...727L..20K,2012ApJ...750...13C,2017ApJ...840...48P}. The presence of gravitational collapse, however, can modify the PDFs to be a power-law distribution in high-density range \citep{1995ApJ...441..702V,2008ApJ...680.1083R,2012ApJ...750...13C,2018ApJ...863..118B,2019MNRAS.482.5233K}. Nevertheless, the effect of radiative transfer potentially can change the PDFs. The correlation of volume density $\rho$ and observed intensity of $I_\nu$ is:
	\begin{equation}
		I_\nu(s)=(I_\nu(0)-S_\nu)e^{-\int nA\phi(\nu)ds}+S_\nu
	\end{equation}
	in which s is the distance along the LOS, $S_\nu$ is the source function, A is the Einstein coefficient, $\phi(\nu)$ is the Doppler broadening function. The molecular gas density $n=x\rho$, where $x$ is the molecular abundance. As a consequence, the role of volume density $\rho$ becomes $e^{-\int xA\phi(\nu)\rho ds}$. The integral term corresponds to the optical depth $\tau_\nu$ for a given molecular specie. High optical depth cancels high-density gas (e.g., $\rm ^{12}CO$ traces volumes density around $\rm 10^{2}$ $\rm cm^{-3}$) and low opacity cancels low-density gas (e.g., $\rm C^{18}O$ traces volumes density around $\rm 10^{4}$ $\rm cm^{-3}$). This cancellation may change the structures or properties of the observed intensity. The corresponding PDFs of molecular intensity is therefore regulated by the volume density and the molecular abundance, i.e., the optical depth. For instance, in the extreme cases of $\tau_\nu\to 0$ or $\tau_\nu\to \infty$, the observed intensity $I_\nu\to I_\nu(0)$ or $I_\nu\to S_\nu$. The PDFs of the integrated intensity are shaped into delta functions.

	In this work, we analyze the PDFs of $\rm ^{12}CO$, $\rm ^{13}CO$, $\rm C^{18}O$ integrated intensity. Here we use the mean optical depth to characterize the PDFs. The higher (or lower) mean optical depth also means the PDF tail’s higher opacity (or the lower optical depth), which means the observed intensity is changed more significantly. We find the PDFs exhibit a power-law tail in the low-intensity range (i.e., positive slope) when the media are optically thick. The power-law tail shifts to the high-intensity range (i.e., negative slope) in the case of extremely optically thin media. This power-law tail is independent of molecular species but only the optical depth. The study by \cite{2019ApJ...881..155P} analytically shows the low-intensity power-law tail can come from the intrinsic properties of turbulent plasma gas. We expect the radiate transfer effect may introduce or enhance these properties.
	
	In terms of the self-gravitating media, the intensity PDFs give limited responses. As shown in Fig.~\ref{fig:pdf_grav} and Fig.~\ref{fig:sk_grav}, the PDFs of $\rm ^{12}CO$, $\rm ^{13}CO$ are insensitive to the self-gravity showing only log-normal distributions. The skewness of $\rm C^{18}O$'s PDFs is positively proportional to the strength of self-gravity. However, $\rm C^{18}O$'s PDFs have difficulties distinguishing the power-law tail induced by the radiative transfer effect and the one induced by self-gravity. Nevertheless, for a given cloud (e.g., a given volume density), a positive skewness indicates either a self-gravitating region or dense structure, and a negative skewness definitely represents a diffuse region, which is crucial for VGT to trace the magnetic fields. In the absence of polarization measurement as a reference, a negative skewness increases the confidence to probe the magnetic fields through VGT. 
	
	The VGT provides an alternative way of identifying the gravitational collapsing regions. According to the MHD turbulence theory \citep{GS95} and turbulent reconnection theory \citep{LV99}, the velocity gradients of turbulent eddies are perpendicular to their local magnetic fields \citep{2000ApJ...539..273C,2002ApJ...564..291C,2001ApJ...554.1175M}. However, this relative orientation of velocity gradients and the magnetic fields becomes parallel in the case of gravitational collapse \citep{YL17b,survey,Hu20}. In the collapsing region boundary, the gradients' orientation exhibits a 90$^\circ$ rapid change. This change can be extracted from the histogram of gradient's orientation over a small sub-region \citep{Hu20}. By scanning the cloud and drawing the histogram, the VGT can identify the gravitational collapsing regions independent of polarimetry measurement. 
	
	Here, we show that the radiative transfer effect does not degrade the performance of VGT. By stacking the gradient maps obtained from multiple emission lines, one can see the POS magnetic fields over different density ranges. The change of velocity gradients also reveals the volume density range in which the collapsing occurs using multiple emission lines. For optically thick tracer $\rm ^{12}CO$, which samples the outskirt diffuse region of the cloud, gives the highest accuracy. However, in observational studies, it was reported that $\rm ^{13}CO$ is more accurate than $\rm ^{12}CO$ comparing with dust polarization \citep{velac,2020arXiv200715344A}. This implies that in molecular gas, the efficiency of dust grain alignment is high with number density $\approx 10^4$ cm$^{-3}$ or greater. Therefore that dust polarization includes the contribution from both intermediate and high-density gas.
	
	In particular, the power-law transition seen in the H$_2$'s PDF can be used to calculate the star formation rate \citep{2005ApJ...630..250K,2011ApJ...743L..29H,2018ApJ...863..118B} and identify the presence of gravitational collapse \citep{1995ApJ...441..702V,2008ApJ...680.1083R,2012ApJ...750...13C,2018ApJ...863..118B,2019MNRAS.482.5233K}. This statistical transition density given by the power-law tail may partially include non-self-gravitating dense gas, as shown in Fig.~\ref{fig:vgt}. Since the velocity gradients are only sensitive to self-gravitating gas, they can confirm the PDFs' identified collapse. The synergy of VGT and the PDFs, i.e., when both approaches are simultaneously used, increases our confidence in the identified collapsing regions, as the example of NGC 1333 used in this work. Also, in the absence of mean magnetic field it is more difficult for the VGT to independently identify the gravitational collapse. The synergy of VGT and PDFs can handle this situation in a similar way.
	
	In VGT's frame, to identify the gravitational collapse, it requires the observational data to cover enough diffuse regions so that the double-peak feature appears (see \S~\ref{sec.theory}). However, this requirement may not be satisfied for a small zoom-in region, such as the NGC 1333 cloud used in this paper. For such a small scale study, once the polarization measurement is available, the comparison of polarization and velocity gradient easily reveals the gravitational collapse. In the lack of dust polarization data, the PDFs assist VGT to probe the magnetic fields. For instance, one can identify the self-gravitating region through PDFs first. Recall that velocity gradients are perpendicular to the magnetic fields in the diffuse region, while gravitational collapse flips their direction by 90$^\circ$ being parallel to the magnetic fields. The magnetic fields can be inferred from {\it re-rotating} the rotated velocity gradients in the self-gravitating region. Note that the density threshold for self-gravitating gas given by the PDFs is a statistical concept. Weak power-law components are likely overwhelmed in the log-normal part or vice versa. Consequently, some self-gravitating gas may be revealed or some non-self-gravitating gas may be included. Nevertheless, these gases only occupy a small fraction which does not significantly degrade the global accuracy of VGT + PDFs in tracing the magnetic field.

	Furthermore, unlike velocity gradients, density gradients flip their direction by 90$^\circ$ in both shocks \citep{YL17b,IGs, H2} and gravitational collapse cases \citep{YL17b,Hu20}. Since the column density PDFs are insensitive to shocks, it can help distinguish the change of density gradients induced by shocks or self-gravity.

	\subsection{Velocity gradients in compressible turbulence}
	The GS95 model considered incompressible MHD (see \S~\ref{sec.theory}), but the in real scenario the ISM turbulence is compressible. By considering the ideal MHD equations and assuming $B_0\gg\delta B$ and the Lagrangian derivative term $\frac{D\delta\boldsymbol{B}}{Dt}\approx0$, where $B_0$ is the mean magnetic field along the z-axis and $\delta B$ is the perturbation components (see Appendix.~\ref{append}), \citet{2020MNRAS.498.1593B} derived the equation:
	\begin{equation}
		\boldsymbol{B_0}(\nabla\cdot\boldsymbol{v})=B_0\partial_z\boldsymbol{v}
	\end{equation}
	which indicates velocity gradients $\partial_z\boldsymbol{v}$ are parallel to the mean magnetic field. In Appendix.~\ref{append}, by adopting the same assumptions, we showed that the displacement vector for the Alfv\'{e}n wave vanishes. It means the parallel relative orientation only appears in compressible turbulence. This agrees with the numerical finding in \citet{LY18a} as well as to the analytical studies of anisotropies in \citet{2012ApJ...747....5L}.  Nevertheless, Alfv\'{e}n modes are the most important for the MHD turbulence \citep{2003MNRAS.345..325C,2007mhet.book...85S}.  Since the wave vector of the Alfv\'{e}nic perturbations in strong turbulence being nearly perpendicular to the local the direction of the magnetic field \citep{2003MNRAS.345..325C}, the anisotropy, and the iso-contours of velocity are both elongated parallel to the magnetic field. The gradients of iso-contours are perpendicular to the local magnetic fields and can be used to trace the magnetic fields \citep{LY18a}, which is the foundation of VGT. In real ISM, the compressible components can contribute parallel relative orientation to VGT. However, their fraction is insignificant and can be suppressed by the sub-block averaging method, as shown in \citet{LY18a}. The compressible modes, therefore, do not degrade the performance of VGT.
	
	\subsection{The intensity PDFs and column density PDFs}
	Three most common ways to derive the column density from observations of molecular line emissions are (i) combining an optically thick line and an optically thin line, for example, $^{12}$CO (1-0) and $^{13}$CO (1-0), assuming that they have the same excitation temperature \citep{2008ApJ...679..481P}, (ii) the curve of growth analysis, which involves using an optically thin line $^{13}$CO (1-0) to estimate the opacity of the optically thick line $^{12}$CO (1-0) \citep{2008ApJ...679..481P}, and (iii) the empirical linear conversion between $^{12}$CO (1-0) integrated intensity and H$_2$ column density, i.e., the so-called X-factor \citep{1987ApJ...319..730S}.
	
	\citet{2009ApJ...692...91G} pointed out that with the combining lines method, there are issues such as abundances and LTE/non-LTE conditions, making the method less reliable than other column density tracers such as dust emission and dust extinction. As \citet{2008ApJ...679..481P} has also shown, empirical linear relations between the integrated intensity and the total column density can likewise perform well in the estimation of column density. Still, each transition is limited on the lower end by the detection threshold and the higher end by the saturation. These empirical linear relations vary substantially from region to region, and thus there is no universal linear relation that applies to all molecular clouds. In any case, these methods all involve assumptions such as a constant abundance, the LTE across the entire cloud, and the opacity/transparency of the molecular line emission. The various physical conditions in different molecular clouds contribute to significant uncertainty of the resulting column density PDF. Therefore, it suffers a substantial uncertainty in deriving an actual column density range column density from emission lines and finding the correlation of intensity PDF and column density PDF.
	
	\subsection{Comparison with earlier work}
	\citet{2013ApJ...771..122B} studied the effects of the radiative transfer on the intensity PDFs of molecular MHD turbulence. They numerically concluded that the integrated intensity maps of $\rm ^{13}CO$ (2–1) transition line with $\tau\simeq1$ have best matching the PDFs of the column density. \citet{2014A&A...564A.106T} later finds that the stellar feedback via ionization can shape the PDFs appearing a low-intensity power-law tail. \cite{2015A&A...575A..79S} discovered the low-intensity power-law tail could also come from the LOS contamination (i.e., noise) on column density structures. This type of power-law tail is observed in the dust extinction data of several Gould Belt molecular clouds  \citep{2019MNRAS.484.3604L}. To exclude the effect of noise, we only use pixels where the brightness temperature is larger than 0.45 K (3$\sigma$ level) to plot the intensity PDFs of NGC 1333 cloud (see \S~\ref{sec:obs}).
	
	In addition, \citet{2016A&A...587A..74S} examined the intensity PDFs of  $^{12}$CO (1-0), $^{13}$CO (1-0), C$^{18}$O (1-0), CS (2-1), and $\rm N_2H^+$ (1-0) for the Cygnus X molecular cloud. A power-law high-intensity tail is observed in the PDFs of later three dense tracers (see Figs.5-6 in \citealt{2016A&A...587A..74S}). They concluded optical depth effects are the main reason for the different PDF shapes. \citet{2016MNRAS.460...82S} then numerically examined the PDFs of $^{12}$CO (1-0), $^{13}$CO (1-0). Their study found a power-law tail in the PDFs' low-intensity range (see the Appendix in \citealt{2016MNRAS.460...82S}). The transition of the PDFs for $\rm ^{12}CO$ column density to a power law at high column density and the association of the power-law part of the gas with the core regions of molecular cloud where self-gravity produces a physical density gradient was shown for particular clouds in M33 in \citet{2018A&A...617A.125C}. Our work extends the study of intensity PDFs to three different CO isotopologs $^{12}$CO (1-0), $^{13}$CO (1-0), C$^{18}$O (1-0), considering the effect of self-gravity. We observe the radiative transfer effect can shape the intensity PDFs independent of self-gravity. In particular, we find in supersonic turbulence the intensity PDFs exhibit (i) a low-intensity power-law tail for the optically thick media $\tau>10$, (ii) a log-normal shape when $0.3<\tau<10$, and (iii) a high-intensity power-law tail for the optically thin media $\tau<0.3$. This change indicates that the skewness goes from positive value to negative value with an increment of $\tau$.

	
	\section{Conclusion} \label{sec:con}
	The VGT and column density PDF have been wildly used to study ISM's magnetic fields and turbulent properties, respectively. They are also useful tools in identifying the self-gravitating gas in non-self-absorbing media. However, in the presence of self-adsorption, molecular tracers sample only parts of the column density structures. The radiative transfer effect can change the properties of intensity PDFs. Therefore, we extend the study of VGT and PDFs, taking into account the presence of self-adsorption in both numerical and observational. To summarize:
	\begin{enumerate}
		\item The accuracy of VGT in tracing the magnetic fields and identifying the self-gravitating regions is insensitive to the radiative transfer effect.
		\item The VGT shows the ability to establish the 3D magnetic field tomography and collapse tomography over a wide range of density using multiple molecular tracers. 
		\item The synergy of VGT and column density PDFs advantageously increases the confidence in identifying gravitational collapsing regions. When the emission line data covers only self-gravitating areas, the synergy could probe the magnetic fields independent of polarization measurement.
		\item For supersonic turbulence, the intensity PDFs of CO isotopologs partially change its shape to a power-law distribution, which effectively depends on the abundance of molecule species or mean optical depth. In particular:
		\begin{enumerate}
			\item The low-intensity power-law tail of intensity PDFs appear in the mean optically thick case that $\tau>10$;
			\item Intensity PDFs are close to log-normal shape when $\tau$ is approximately in the range $0.3\le\tau\le10$;
			\item The high-intensity power-law tail of intensity PDFs appear in the case of $\tau<0.3$. 
			\item This power-law feature in the intensity PDFs can be independent of self-gravity 
		\end{enumerate}
		\item The observed shape of the intensity PDFs is strongly affected by self-absorption. This can strongly interfere with the attempts to use the intensity PDFs for identifying gravitational collapsing regions in molecular clouds. 
		\item A negative skewness of intensity PDF appears in optically thick diffuse media. This skewness indicates a higher accuracy of VGT in probing the magnetic fields.
		\item We confirm the collapsing material occupies a large fraction in the NGC 1333 molecular cloud.
	\end{enumerate}
	
	\section*{Acknowledgements}
	We thank Ka Wai Ho and Ka Ho Yuen for helpful discussions. A.L. acknowledges the support of the NSF grant AST 1715754, and 1816234. Y.H. acknowledges the support of the NASA TCAN 144AAG1967. Flatiron Institute is supported by the Simons Foundation. The work uses observations obtained with Planck (http://www.esa.int/Planck), an ESA science mission with instruments and contributions directly funded by ESA Member States, NASA, and Canada.  The HGBS is a Herschel Key Programme jointly carried out by SPIRE Specialist Astronomy Group 3 (SAG 3), scientists of several institutes in the PACS Consortium (CEA Saclay, INAF-IFSI Rome and INAF-Arcetri, KU Leuven, MPIA Heidelberg), and scientists of the Herschel Science Center (HSC).
	\section*{Data availability}
	The data underlying this article will be shared on reasonable request to the corresponding author.
	
	\appendix
	\section{Velocity gradients in compressible turbulence}
	\label{append}
	Following the steps used in \citet{2020MNRAS.498.1593B}, firstly we have the ideal MHD equations:
	\begin{equation}
		\begin{aligned}
			\frac{\partial\rho}{\partial t}+\nabla\cdot\rho\boldsymbol{v} &=0\\
			(\frac{\partial}{\partial t}+\boldsymbol{v}\cdot\nabla)\rho\boldsymbol{v}&=\frac{(\boldsymbol{B}\cdot\nabla)\boldsymbol{B}}{4\pi}-\nabla(c_s^2\rho+\frac{\boldsymbol{\vert B\vert}^2}{8\pi})+\rho\boldsymbol{F}\\
			\frac{\partial B}{\partial t}&=\nabla\times (\boldsymbol{v} \times\boldsymbol{B})\\
			\nabla\cdot \boldsymbol{B}&=0
		\end{aligned}
	\end{equation}
	where $\boldsymbol{v}$ is the fluid velocity, $\rho$ is the fluid density, $\boldsymbol{B}$ is the magnetic field, $c_s$ is the sound speed and $\boldsymbol{F}$ is an external driving force.
	By considering a magnetic field in the form of:
	$$
	\begin{pmatrix} 
		B_x \\
		B_y \\
		B_z
	\end{pmatrix}
	=\begin{pmatrix} 
		0 \\
		0 \\
		B_0
	\end{pmatrix}+
	\begin{pmatrix} 
		\delta B_x \\
		\delta B_y \\
		\delta B_z
	\end{pmatrix}
	$$
	,  \citet{2020MNRAS.498.1593B} simplify the equation to:
	\begin{equation}
		(\frac{\partial}{\partial t}+ \boldsymbol{v}\cdot\nabla)\delta \boldsymbol{B}=B_0\partial_z\boldsymbol{v}+(\delta\boldsymbol{B}\cdot\nabla)\boldsymbol{v}-(\boldsymbol{B_0}+\delta\boldsymbol{B})(\nabla\cdot \boldsymbol{v})
	\end{equation}
	
	In sub-Alfv\'{e}nic molecular clouds, we can assume $B_0\gg\delta B$ and the Lagrangian derivative term $\frac{D\delta\boldsymbol{B}}{Dt}\approx0$. \citet{2020MNRAS.498.1593B} therefore obtained the equation:
	\begin{equation}
		\label{eq:vb}
		\boldsymbol{B_0}(\nabla\cdot\boldsymbol{v})=B_0\partial_z\boldsymbol{v}
	\end{equation}
	which indicates velocity gradients are parallel to the mean magnetic field. This consideration is drawn in the frame of compressible turbulence. It therefore naturally to explore which MHD mode (Alfv\'{e}n, fast, and slow modes) contribute to this parallel relative orientation. Here we follow the mode decomposition procedure used in \citet{2003MNRAS.345..325C}, which assumes the displacements $\boldsymbol{\xi}(\boldsymbol{r},t)$ of which $\partial\boldsymbol{\xi}(\boldsymbol{r},t)/\partial t=\boldsymbol{v}$ vanish at t = 0, the momentum equation is simplified to:
	\begin{equation}
		\label{eq.A4}
		\begin{aligned}
			\ddot{\boldsymbol{\xi}}/v_A^2-\nabla[(\alpha+1)\nabla\cdot\boldsymbol{\xi}-(\hat{\boldsymbol{B_0}}\cdot \nabla)(\hat{\boldsymbol{B_0}}\cdot \boldsymbol{\xi})]\\
			-(\hat{\boldsymbol{B_0}}\cdot \nabla)^2\boldsymbol{\xi}+(\hat{\boldsymbol{B_0}}\cdot \nabla)(\nabla\cdot\boldsymbol{\xi})\hat{\boldsymbol{B_0}}=0
		\end{aligned}
	\end{equation}
	here $\alpha=c_s^2/v_A^2$ and $v_A=B_0/\sqrt{4\pi\rho}$. By adopting the Eq.~\ref{eq:vb}, the last two terms in Eq.~\ref{eq.A4} vanish. In Fourier space Eq.~\ref{eq.A4} becomes:
	\begin{equation}
		\ddot{\boldsymbol{\xi}}/v_A^2+k\hat{\boldsymbol{k}}[(\alpha+1)k\xi_k-k_\parallel\xi_\parallel]=0
	\end{equation}
	where $\xi_k=\boldsymbol{\xi}\cdot\hat{\boldsymbol{k}}$, $\xi_\parallel=\boldsymbol{\xi}\cdot\hat{\boldsymbol{k}}_\parallel$, $\hat{\boldsymbol{k}}=\boldsymbol{k}/k$, and $\hat{\boldsymbol{k}}_\parallel$ is a unit vector parallel to $\boldsymbol{B_0}$. Assuming $\ddot{\boldsymbol{\xi}}=-\omega^2\boldsymbol{\xi}=-c^2k^2\boldsymbol{\xi}$, we can rewrite the above equation as:
	\begin{equation}
		\boldsymbol{\xi}c^2/v_A^2-\hat{\boldsymbol{k}}[(\alpha+1)\xi_k-\cos\theta\xi_\parallel]=0
	\end{equation}
	where $\cos\theta=k_\parallel/k$ is the angle between $\boldsymbol{k}$ and $\boldsymbol{B_0}$. Then Using $\hat{\boldsymbol{k}}=\sin\theta\hat{\boldsymbol{k}}_\bot+\cos\theta\hat{\boldsymbol{k}}_\parallel$:
	\begin{equation}
		\boldsymbol{\xi}c^2/v_A^2-\sin\theta\hat{\boldsymbol{k}}_\bot[(\alpha+1)\xi_k-\cos\theta\xi_\parallel]-\cos\theta\hat{\boldsymbol{k}}_\parallel[(\alpha+1)\xi_k-\cos\theta\xi_\parallel]=0
	\end{equation}
	Writing $\boldsymbol{\xi}=\xi_\bot\hat{\boldsymbol{k}}_\bot+\xi_\parallel\hat{\boldsymbol{k}}_\parallel+\xi_\phi\hat{\boldsymbol{k}}_\phi$, we get three equations:
	\begin{equation}
		\label{eq.A8}
		c^2/v_A^2\xi_\bot-\sin\theta[(\alpha+1)\xi_k-\cos\theta\xi_\parallel]=0
	\end{equation}
	\begin{equation}
		\label{eq.A9}
		c^2/v_A^2\xi_\parallel-\cos\theta[(\alpha+1)\xi_k-\cos\theta\xi_\parallel]=0\end{equation}
	\begin{equation}
		\label{eq.A10}
		c^2/v_A^2\xi_\phi=0
	\end{equation}
	there is only a trivial solution for Eq.~\ref{eq.A10}, which corresponds to the Alfv\'{e}n mode since the direction of the
	displacement vector for the Alfv\'{e}n wave is parallel to the azimuthal basis \citep{2003MNRAS.345..325C}. The parallel relative orientation suggested by Eq.~\ref{eq.A4} \citep{2020MNRAS.498.1593B}, therefore, does not hold for the incompressible MHD turbulence. The solutions of Eq.~\ref{eq.A8} and Eq.~\ref{eq.A9} can be obtained by using: $\xi_k=\xi_\parallel\cos\theta+\xi_\bot\sin\theta$. After rearrangement we have:
	\begin{equation}
		[\frac{c^2}{v_A^2}-(\alpha+1)\sin^2\theta]\xi_\bot-\alpha\cos\theta\sin\theta\xi_\parallel=0\\
	\end{equation}
	\begin{equation}
		[\frac{c^2}{v_A^2}-\alpha\cos^2\theta]\xi_\parallel-(\alpha+1)\cos\theta\sin\theta\xi_\bot=0 
	\end{equation}
	Combining these two, we get:
	\begin{equation}
		[\frac{c^2}{v_A^2}-\alpha\cos^2\theta][\frac{c^2}{v_A^2}-(\alpha+1)\sin\theta^2]=\alpha(\alpha+1)\cos^2\theta\sin^2\theta
	\end{equation}
	The roots of the equation are:
	\begin{equation}
		\label{eq.A14}
		c^2=v_A^2(\alpha+\sin^2\theta)
	\end{equation}
	Comparing wit the solutions of fast, slow mode \citep{2003MNRAS.345..325C}: 
	\begin{equation}
		c_{f,s}^2=\frac{1}{2}v_A^2[(1+\alpha)\pm\sqrt{(1+\alpha)^2-4\alpha\cos^2\theta}]
	\end{equation}
	where the subscripts ‘f’ and ‘s’ stand for ‘fast’ and ’slow’ waves, respectively, we find when $\theta=0$ and $\rm \alpha=(M_A/M_S)^2=\frac{1}{2}\beta<1$, Eq.~\ref{eq.A14} corresponds to the pure slow mode, while $\theta=0$ and $\rm \alpha>1$ is the pure fast mode. In the case of $\theta=90^\circ$, Eq.~\ref{eq.A14} represents the pure fast mode. 
	

	
	
\end{document}